\documentclass[aps,prl,reprint,superscriptaddress,showpacs,preprintnumbers,amsmath,amssymb]{revtex4-2}
\usepackage[dvipdfmx]{graphicx}

\usepackage[breaklinks,colorlinks,citecolor=blue,urlcolor=blue]{hyperref}
\usepackage{bm}        
\usepackage{amssymb, amsmath, amsfonts}   
\usepackage{color}
\usepackage[normalem]{ulem}
\DeclareGraphicsExtensions{{.pdf}}
\begin{document}
\newcommand{\noter}[1]{{\color{red}{#1}}}
\newcommand{\noteb}[1]{{\color{blue}{#1}}}
\newcommand{\field}{\left( \boldsymbol{r}\right)}
\newcommand{\paren}[1]{\left({#1}\right)}
\newcommand{\vect}[1]{\boldsymbol{#1}}
\newcommand{\uvect}[1]{\tilde{\boldsymbol{#1}}}
\newcommand{\vdot}[1]{\dot{\boldsymbol{#1}}}
\newcommand{\vder}{\boldsymbol{\nabla}}

\renewcommand{\phi}{\varphi}
\newcommand{\phiJ}{\varphi_{\rm J}}
\newcommand{\be}{\begin{equation}}
\newcommand{\ee}{\end{equation}}
\newcommand{\bea}{\begin{equnaray}}
\newcommand{\eea}{\end{equnaray}}
\newcommand{\ba}{\begin{align}}
\newcommand{\ea}{\end{align}}
\newcommand{\ave}[1]{\left\langle {#1} \right\rangle}
\newcommand{\tk}[1]{{\color{red}{#1}}}
\newcommand{\TK}[1]{{\color{blue}{#1}}}
%
\widetext
%
%
\title{Probing Anharmonic and Heterogeneous Carrier Dynamics Across Sublattice Melting in a Minimal Model Superionic Conductor}

\author{Sucharita Niyogi}
\email{niyogi.sucharita.d3c``@"osaka-u.ac.jp}
\affiliation{D3 Center, The University of Osaka, Toyonaka 560--0043, Japan}
\affiliation{Department of Physics, The University of Osaka, Osaka 560--0043, Japan}
\affiliation{Department of Physics, Nagoya University, Nagoya 464--8601, Japan}

\author{Takenobu Nakamura}
\affiliation{National Institute of Advanced Industrial Science and Technology, Ibaraki 305--8568, Japan}

\author{Genki Kobayashi}
\affiliation{Solid State Chemistry Laboratory, RIKEN, Saitama 351--0198, Japan}

\author{Yasunobu Ando}
\affiliation{Institute of Integrated Research, Institute of Science Tokyo, Kanagawa 226--8501, Japan}

\author{Takeshi Kawasaki}
\email{kawasaki.takeshi.d3c``@"osaka-u.ac.jp}
\affiliation{D3 Center, The University of Osaka, Toyonaka 560--0043, Japan}
\affiliation{Department of Physics, The University of Osaka, Osaka 560--0043, Japan}
\affiliation{Department of Physics, Nagoya University, Nagoya 464--8601, Japan}
\date{\today}
\begin{abstract}
Despite decades of research, the microscopic origin of sublattice melting and fast ion transport in superionic conductors remains elusive. Here, we introduce a chemically neutral \emph{minimal binary model} consisting of a rigid host lattice stabilized by short-range steric repulsion and a soft carrier sublattice interacting via long-range Wigner-type forces. This contrast naturally produces distinct melting temperatures and an intermediate \emph{sublattice-melting} phase in which carriers become fluidlike while the host remains crystalline. Molecular dynamics simulations identify three dynamical regimes-crystalline, sublattice-melt, and fully molten-marked by sharp changes in diffusivity, structural correlations, and dynamical heterogeneity. Near sublattice melting, carrier motion is strongly \emph{anharmonic} and spatially heterogeneous, beyond mean-field hopping descriptions. By tuning the density, we demonstrate that sublattice melting can be continuously controlled, establishing a direct link between lattice softness, anharmonicity, and collective ion transport. Comparison with conventional long-range Coulombic models confirms that our minimal model reproduces the key dynamical signatures of superionicity, providing a unified microscopic foundation for designing mechanically robust superionic conductors.
\end{abstract}
%
\maketitle
%
\section{Introduction}
The rapid advancement of energy storage technologies has placed rechargeable batteries at the center of modern society, powering applications from portable electronics to electric vehicles and large-scale renewable energy integration. At the core of all such battery technologies is ionic transport: ionic conductors, whether liquid or solid, are essential for efficient charge transport and directly determine battery performance, longevity, and safety. While liquid electrolytes in conventional Li-ion batteries provide high ionic mobility and performance  \cite{Denholm2023}, these volatile solvents are flammable, chemically reactive, and prone to leakage, making them a primary source of thermal runaway and catastrophic failure in high-energy-density devices \cite{KongEnergies2018}. Therefore, overcoming these limitations is crucial for the safe and reliable deployment of next-generation batteries.

In this context, solid-state electrolytes, particularly superionic conductors \cite{CavazzoniScience1999, HernandezPRL2016, LinPRE2020, LinPNAS2023, RichardsNatComm2016, WangNatMat2015, YajimaJMCA2021, BoyceSSC1977b, OwensScience1967, KannoMRB1991, DenesSSI1996, KamayaNatMater2011, HayashiNatCommun2019, TakeiriNatMater2022, HullRPP2004, FunkeSTAM2013, YamamotoSTAM2017,VerbraekenNatMatter2015}, constitute a distinct class of materials in which ions move almost as freely as in molten salts while the host lattice retains long-range crystalline order. Their defining feature is the coexistence of two distinct sublattices: an immobile framework that maintains structural integrity and a mobile-ion sublattice that becomes disordered at elevated temperatures \cite{RettieNatMater2021}. 
The temperature dependence of this order-disorder transition varies widely among materials. For example, NaCl exhibits a pronounced increase in ionic conductivity only near its melting point, consistent with conventional melting behavior, whereas some superionic conductors, such as PbF$_2$, show a gradual enhancement of conductivity over several hundred Kelvin, well below the melting temperature \cite{Kawamura2017, HullRPP2004, FunkeSTAM2013}. Others, including AgI, undergo a first-order phase transition accompanied by an abrupt increase in ionic conductivity by more than three orders of magnitude.
Similar melting-like transport behavior has also been reported in experiments on other charge carriers, including hydride ions (H$^{-}$), indicating the generality of this phenomenon across some ion species \cite{TakeiriNatMater2022,VerbraekenNatMatter2015}. \noindent Despite these observations, the microscopic origin of such an anomalous transition remains unresolved. Some recent analyses of layered and tunnel-type superionic conductors have shown that carrier transport is strongly influenced by many-body interactions among mobile ions, reshaping the diffusion landscape beyond simple single-particle hopping; here, Raman measurements and MD simulations further show that short-range cation-cation repulsion modifies activation barriers, underscoring the collective nature of fast-ion motion \cite{KamishimaJPSJ2010}. Together, these studies suggest that \emph{sublattice melting} and strong \emph{anharmonic} lattice fluctuations are both essential ingredients for superionic transport; however, a direct microscopic link connecting these phenomena remains to be clearly established.

To place these observations in context, \emph{sublattice melting} \cite{Hainovsky1995, LinPRE2020, WelchJEM1975, BoyceSSC1977a, BoyceSSC1977b, IshiiJPSJ1998} refers to a state in which the mobile ionic network becomes dynamically disordered while the host framework remains crystalline, enabling liquid-like conductivity without structural collapse. Early mean-field models \cite{WelchJEM1975} attributed this transition to a balance between defect formation energy and configurational entropy, predicting disorder within one ionic sublattice prior to complete melting. However, their defect-based mean-field framework precludes access to real-space ion dynamics, spatial correlations, and anharmonic lattice effects central to superionic transport. Experimental studies on CuI provided early dynamical evidence for sublattice melting: NMR measurements showed anomalously enhanced relaxation near the superionic transition, consistent with a molten copper sublattice within an ordered iodide framework \cite{BoyceSSC1977a, BoyceSSC1977b}. Subsequent theoretical treatments \cite{Hainovsky1995} described the precursor defect proliferation using a cube-root dependence of the defect chemical potential, reproducing the premelting characteristics observed in AgI and PbF$_2$.
More recently, analogous phenomena have been realized in asymmetric colloidal crystals \cite{LinPRE2020}, where smaller charged particles delocalize within a crystalline matrix of larger oppositely charged spheres, and in ultrasmall copper selenide clusters exhibiting liquid-like cationic sublattices even at ambient temperature \cite{WhiteNatComm2017}. Complementary experimental studies have since directly visualized this selective sublattice melting \cite{DingNatPhys2025}, reaffirming its universality in ion-conducting solids. Despite these advances, the microscopic origin of such selective disorder remains poorly understood, as most existing descriptions rely on mean-field or defect-based frameworks that neglect correlated particle motion and its feedback on the lattice vibrational landscape.
An instructive analogy is provided by colloidal Wigner crystals, where crystalline order is maintained despite large-amplitude particle fluctuations due to significant configurational entropy \cite{SprakelPRL2017}, highlighting the role of anharmonic dynamics within an ordered framework.

Selective sublattice disorder reshapes the vibrational potential energy landscape, introducing \emph{anharmonicity} that enables nonlinear energy exchange and mode coupling between phonons \cite{HoshinoSSC1977, WolfJGR1984, BrennerJPCL2022}. Such anharmonicity effectively softens the lattice and lowers local activation barriers for ionic motion. Early studies on superionic AgI \cite{HoshinoSSC1977} proposed that Ag$^{+}$ ions undergo strongly anharmonic, semi-liquid-like thermal vibrations that extend toward neighboring interstitial sites. This anharmonic motion can overcome local potential barriers and promotes ionic hopping, while appearing in diffraction as a pseudo-static occupation of interstitial sites rather than true site disorder. High-pressure analyses \cite{WolfJGR1984} further showed that harmonic or mean-field descriptions fail to capture melting behavior driven by such vibrations. More recent first-principles and spectroscopic studies on sodium-ion conductors \cite{BrennerJPCL2022} demonstrated that strong host-ion anharmonic coupling induces order-disorder transitions with soft modes persisting across the phase change, directly linking lattice dynamics to fast ionic conduction. Nevertheless, establishing a unified microscopic description that connects anharmonic lattice dynamics with emergent collective transport behavior remains an important challenge.

Together, these studies suggest that sublattice melting and anharmonicity are not independent phenomena but mutually reinforcing aspects of the same underlying physics, in which selective ionic disorder feeds back into lattice softening, and vice versa. In particular, recent work on Li$_{10}$GeP$_2$S$_{12}$ has shown that Li$^+$ conduction proceeds via correlated and cooperative migration of densely packed ions \cite{YajimaJMCA2021}, explicitly invoking the concept of \emph{dynamical heterogeneity} through the coexistence of fluid-like, highly mobile regions and immobile rigid domains. 
This coexistence of mobile and immobile regions is a well-known hallmark of glass-forming systems and other dynamically heterogeneous materials.
More broadly, dynamical heterogeneity is a central concept in glass-forming systems, where it underlies transport anomalies and cooperative motion \cite{KobPRL1997, YamamotoJPS1997, YamamotoPRE1998, BerthierScience2005, DauchotPRL2005, KawasakiPRL2007, BerthierRevModPhys2011, KawasakiSciAdv2017}. At the microscopic level, such heterogeneous dynamics have been linked to localized anharmonic vibrational fluctuations and soft quasilocalized nonphononic modes that promote collective rearrangements \cite{ZylbergPNAS2017, TanguyEPL2010}.
In this light, dynamical heterogeneity offers a natural conceptual bridge between sublattice melting and anharmonicity in superionic conductors: spatial variations in local anharmonicity amplify mobility contrasts, giving rise to cooperative, string-like dynamics beyond mean-field descriptions, as observed in glass-forming liquids where enhanced vibrational fluctuations precede collective string-like rearrangements \cite{KawasakiJCP2013}. The extent to which analogous vibrational-heterogeneity couplings govern fast-ion transport in crystalline solids remains an open question.


To elucidate this connection, we employ a minimal and tractable model to systematically probe how temperature and interactions control sublattice melting and collective carrier motion. This approach allows us to identify the conditions under which liquid-like regions emerge within an otherwise rigid lattice, quantify correlated ion dynamics beyond mean-field descriptions, and explore the emergence of superionic conduction. By linking lattice softness, anharmonicity, and dynamical heterogeneity, this framework provides a simplified route to understand fast-ion transport and design mechanically robust, high-performance superionic conductors.
\section{Results}
Our analysis focuses on three key mechanisms-selective sublattice melting, structural heterogeneity, and anharmonicity-to elucidate ion-transport dynamics in two-dimensional systems. In two dimensions (2D), true long-range translational order is prohibited at finite temperature by the Hohenberg-Mermin-Wagner theorem \cite{MerminWagnerPRL1966,FlennerNatCommn2015,ShibaPRL2016, ShibaJPCM2018, ShibaPRL2019,FlennerPNAS2019}, although the orientational order may persist. Our conclusions do not rely on strict translational long-range order; rather, they concern the emergence of liquid-like transport within an ordered host through enhanced anharmonicity and spatially heterogeneous, string-like motion. 2D systems are therefore used primarily for clarity and visualization, while corresponding three-dimensional (3D) results are presented in the Supporting Information.

\subsection{Microscopic configurations and sublattice disorder}
To investigate the microscopic dynamics of the 2D host-carrier system, we employ our non-additive potential (NAP) model systems in which the host and carrier form stable crystals at low temperatures.
The choice of such a system reflects the long-range nature of the interaction potential, analogous to the Coulomb force (see ``Materials and Methods'' for details).
Under these conditions, low-density carriers naturally organize into ordered patterns at low temperatures, forming a well-defined crystalline state. As the temperature increases, the host lattice remains crystalline while the carrier sublattice progressively melts.
This configuration provides a mechanically stable reference state from which the subsequent thermal evolution and melting processes can be systematically explored, ensuring that the high-temperature dynamics originate from physically meaningful particle arrangements established at low temperatures.
\begin{figure}[!t]
    \centering
    \includegraphics[width=\linewidth]{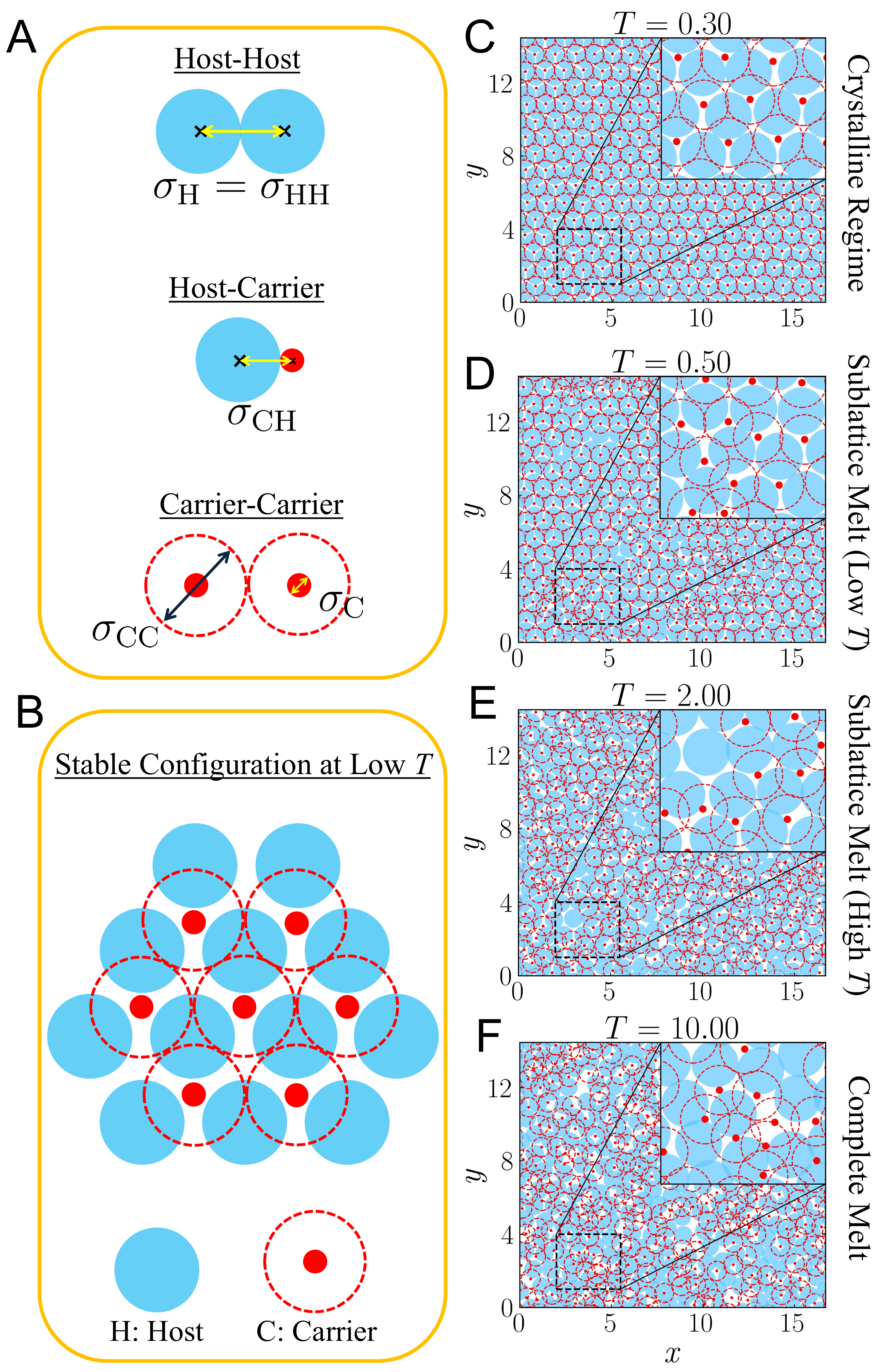}
    \caption{ 
    \textbf{Structural representations of the host-carrier in 2D NAP model:} (\textit{A}) Schematic representation of the interaction scheme between interparticle pairs: the \emph{host-host} interaction length is denoted as $\sigma_\mathrm{H} = \sigma_\mathrm{HH}$, the \emph{host-carrier} interaction length as $\sigma_\mathrm{CH}$, the non-additive interaction length $\sigma_\mathrm{CC}$ between \emph{carrier-carrier} is represented by the red dotted line, and the \emph{effective-carrier} interaction length is chosen to be $\sigma_\mathrm{C}$ (see ``Materials and Methods'' for more details). (\textit{B}) Low-temperature configuration obtained using a Wigner-type interaction, illustrating the ordered host lattice (blue) with carrier particles (red) occupying energetically favorable interstitial sites. (\textit{C--F}) Representative snapshot configurations illustrating crystalline, sublattice-melt, and fully molten states across increasing temperature (top to bottom) for area packing fraction $\phi$ = 0.85. For clarity, each snapshot includes an inset highlighting a magnified region of the configuration, emphasizing local structural arrangements and carrier environments.}
    \label{fig:fig1}
\end{figure}
\begin{figure*}[t]
    \centering
    \includegraphics[width=\linewidth]{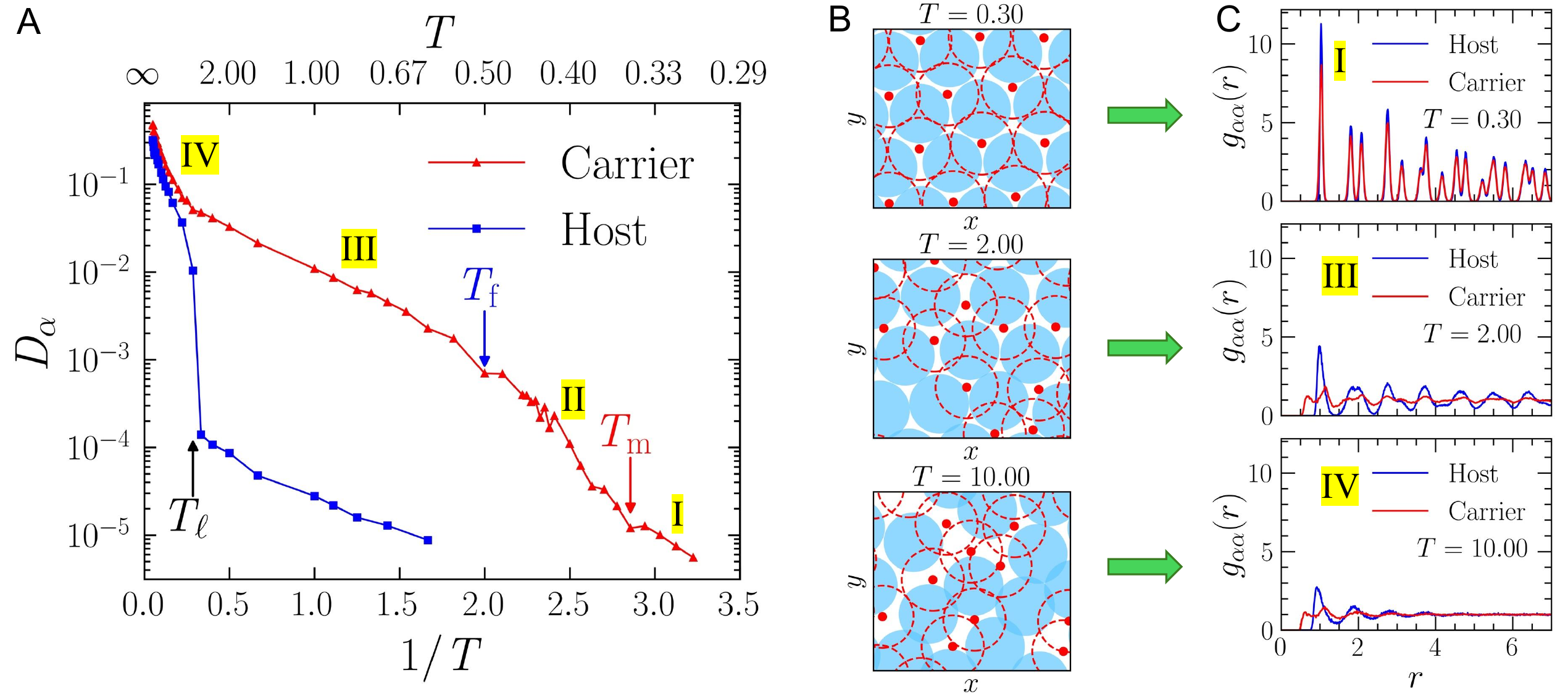}
    \caption{%
        \textbf{Selective sublattice melting and structural evolution in 2D NAP model for $\phi = 0.85$:} (\textit{A}) Schematic representation of the diffusivity $D_{\alpha}$ ($\alpha \in \{{\rm C, H}\}$, where $C$: Carrier and $H$: Host) as a function of inverse temperature $1/T$, illustrating four distinct regimes: (I) crystalline, (II) onset of sublattice melting, (III) sublattice melting, and (IV) full melting. Three characteristic temperatures $T_\mathrm{m} (\sim 0.35)$, $T_\mathrm{f} (\sim 0.50)$, and $T_\mathrm{\ell} (\sim 3.50)$ are indicated, corresponding to the onset of sublattice melting, the freezing of carrier motion, and the liquid-like regime, respectively. (\textit{B}) Representative snapshots illustrating structural evolution across these regimes. (\textit{C}) Radial distribution functions $g_{\alpha\alpha}(r)$ corresponding to each regime. Region III shows disordered carrier sublattice coexisting with ordered host lattice, evidencing selective carrier melting.
    }
    \label{fig:fig2}
\end{figure*}
Figure~\ref{fig:fig1} illustrates the structural framework underpinning our 2D~NAP model. The host and carrier particles interact through non-additive pair potentials that define distinct effective length scales, as summarized in Fig.~\ref{fig:fig1}\textit{A}. These interaction parameters are coarse-grained, effective quantities that encode collective lattice stiffness and carrier confinement, rather than specific microscopic chemistry. At low temperature (Fig.~\ref{fig:fig1}\textit{B}), the interactions stabilize a crystalline arrangement in which carriers occupy well-defined interstitial positions within the host lattice-a reference configuration from which all subsequent temperature-dependent transformations are analysed. To provide a broader picture, Figs.~\ref{fig:fig1}\textit{C--F} present representative structural snapshots at different temperatures, where one can clearly observe the progressive loss of order. In the most stable, low-temperature state ($T = 0.30$ in Fig.~\ref{fig:fig1}\textit{C}; for units see ``Materials and Methods''), both host and carrier particles exhibit a well-preserved crystalline symmetry. Upon heating, a sublattice melting transition emerges (Figs.~\ref{fig:fig1}\textit{D,E}). At the lower bound of this regime ($T = 0.50$), partial disordering of the carrier sublattice is observed-regions of crystalline order coexist with melted domains-while the host lattice remains intact. At a higher temperature ($T = 2.00$), the carrier sublattice becomes fully disordered, yet the host lattice preserves its crystalline arrangement. This indicates that both states lie within the sublattice melting regime where only the carrier network loses structural order. At sufficiently high temperatures ($T = 10.00$ in Fig.~\ref{fig:fig1}\textit{F}), both host and carrier particles lose positional order, corresponding to complete melting of the system. 

To quantify particle mobility and capture the key characteristics underlying this phenomenon, we computed the mean-squared displacement (MSD) over time of the carrier and host particles separately (see ``Materials and Methods'' and Fig.~\ref{fig:fig4}\textit{A}). The long-time diffusive behavior of the MSD was then used to calculate the self-diffusion coefficient $D_{\alpha}$ ($\alpha \in \{{\rm C, H}\}$, where $C$: Carrier and $H$: Host) using the Einstein relation (see ``Materials and Methods''). This analysis was performed separately for both carrier and host particles, allowing us to directly compare their respective mobilities and the influence of temperature on their transport dynamics.
In Fig.~\ref{fig:fig2}\textit{A}, we identify four distinct dynamical regimes characterized by changes in the temperature dependence of the diffusivity. Importantly, these regimes are not introduced as empirical classifications, but emerge from a progressive change in the underlying microscopic dynamics.
At low temperatures (regime I), both host and carrier particles exhibit nearly harmonic vibrations around their equilibrium positions. In this regime, particle motion is dominated by independent, thermally activated hopping over well-defined local energy barriers, leading to Arrhenius behavior with comparable slopes for both species.
As the temperature increases toward the onset of sublattice melting (regime II), anharmonic lattice fluctuations become significant. These fluctuations locally distort the confining potential experienced by the carriers, effectively softening the energy landscape and reducing the activation barriers for migration. As a result, particle motion begins to deviate from independent hopping and increasingly involves correlated displacements.
In the sublattice-melting regime (regime III), these effects become dominant. The enhanced anharmonicity leads to the formation of spatially heterogeneous regions with reduced local stiffness, which act as pathways for cooperative carrier motion. Transport is therefore no longer governed by a single-particle activation process, but by collective, string-like rearrangements involving multiple carriers. This crossover is reflected in the marked enhancement of carrier diffusivity and the breakdown of a simple Arrhenius description.
At even higher temperatures (regime IV), once the host lattice loses structural integrity, the system transitions to a fully molten state. In this regime, both species exhibit homogeneous, liquid-like dynamics, and transport is governed by standard kinetic mechanisms rather than by cooperative processes within a solid framework.
Taken together, these results demonstrate that the temperature-dependent transport behavior arises from a continuous evolution of the energy landscape: increasing anharmonicity reduces effective activation barriers and promotes cooperative dynamics, ultimately leading to sublattice melting and liquid-like transport.
These findings are in good agreement with previous experimental observations that reported similar melting signatures in related systems~\cite{Ding2020, TakeiriNatMater2022,VerbraekenNatMatter2015}. A direct comparison between our 2D~NAP model predictions and available experimental trends is provided in the Supporting Information.

To aid the interpretation of the transport regimes, Fig.~\ref{fig:fig2}\textit{B} shows magnified cross-sectional views at representative temperatures ($T = 0.30$, $2.00$, and $10.00$), corresponding to the crystalline, sublattice-melted, and fully molten states already discussed in Figs.~\ref{fig:fig1}\textit{C, E, F}. These views highlight the progressive loss of order in the carrier sublattice while the host lattice remains intact in the sublattice-melting regime, followed by complete disordering of both components at high temperature.

To quantitatively characterize these structural correlations, we calculate the radial distribution function (RDF), $g_{\alpha\alpha}(r)$, for both carrier and host particles (see ``Materials and Methods''). As shown in Fig.~\ref{fig:fig2}\textit{C}, the RDF exhibits sharp periodic peaks at low temperature (region I), reflecting the long-range crystalline order of both the host and the carrier particles. In the intermediate regime (region III), the contrast becomes evident: the RDF of the host particles retains pronounced peaks even at higher temperatures (e.g., $T = 2.00$), while that of the carriers broadens into liquid-like correlations, corroborating the sublattice melting identified from the snapshots. At very high temperatures (e.g., $T = 10.00$ in region IV), both sublattices lose positional order, and the RDFs of hosts and carriers display the characteristic liquid-like structure with damped oscillations. A detailed analysis of the temperature evolution of the carrier and host RDFs, including the emergence of excess short-range features associated with activated carrier hopping, is provided in the Supporting Information.
\subsection{Dynamical aspect of sublattice melt}
\begin{figure}[t]
    \centering
    \includegraphics[width=\linewidth]{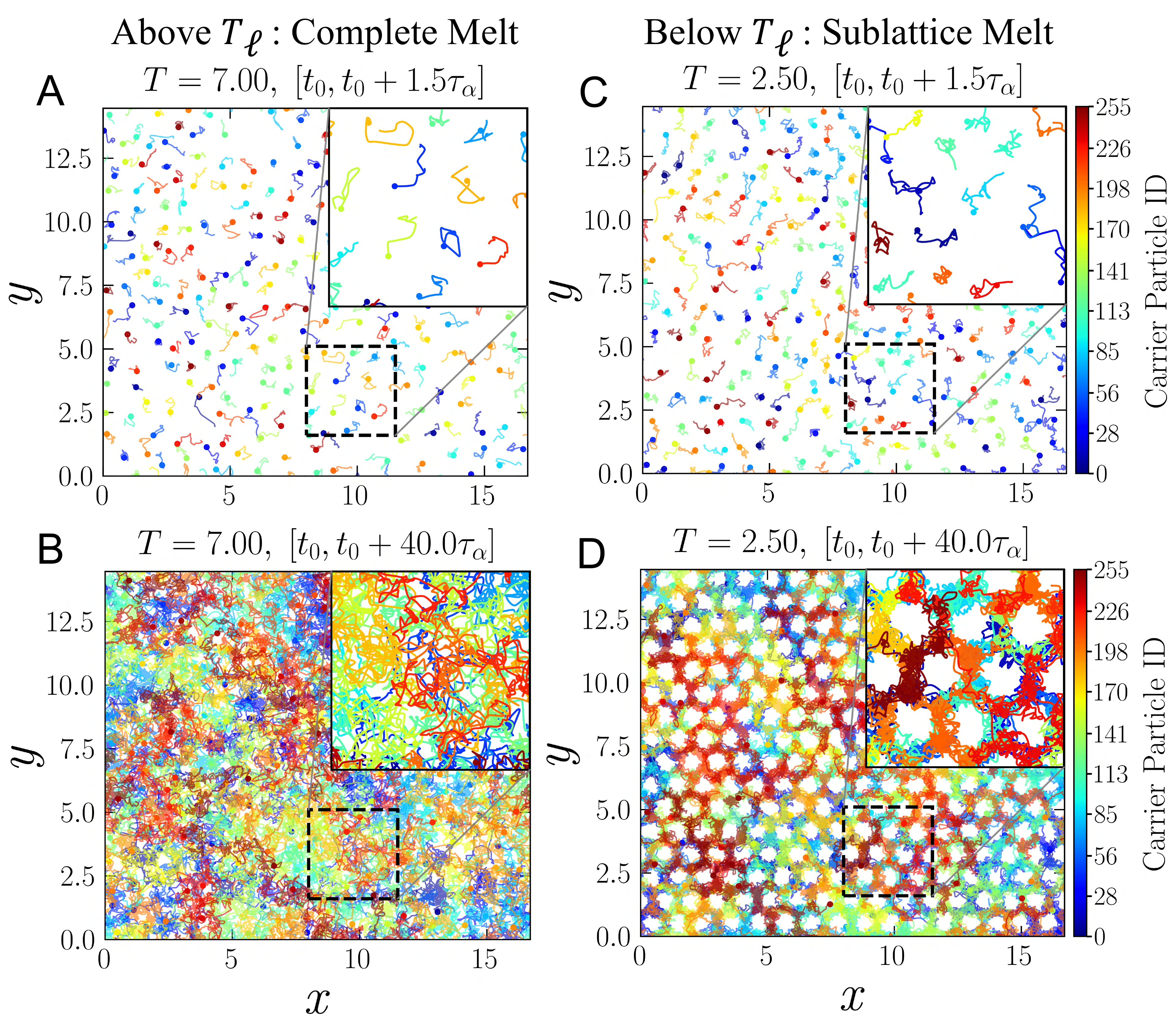}
    \caption{%
        \textbf{Trajectory of carriers at different time intervals and temperature regimes for our 2D NAP model for $\phi = 0.85$:} Panels (\textit{A--D}) show carrier trajectories within region III-IV of the diffusivity plot (as shown in Fig.~\ref{fig:fig2}\textit{A}), corresponding to the sublattice melting regime. Panels (\textit{A, B}) present trajectories above the melting temperature of host within region IV, beyond the sublattice melting regime, while panels (\textit{C, D}) correspond to the higher-temperature side of the sublattice melting regime (region III). Panels (\textit{A, C}) correspond to shorter observation time windows, whereas panels (\textit{B, D}) show the dynamics over longer time windows. Carrier particle indices (ID) are color-coded as indicated by the colorbar on the right-hand side. For corresponding movies of these trajectories, check the Supporting Information.
    }
    \label{fig:fig3}
\end{figure}

While the previous analyses characterize the structural and transport properties of the system through static configurations and diffusivity measurements, a direct visualization of particle trajectories provides additional insight into the dynamical aspects of sublattice melting. To this end, we trace the trajectories of representative carrier particles within the host lattice and map their motion across distinct time windows and temperature regimes. Figure~\ref{fig:fig3} presents the carrier trajectories at representative temperatures for area packing fraction $\phi = 0.85$, illustrating the evolution of carrier dynamics across the sublattice melting transition.
At high temperatures, e.g.~$T = 7.00$ (see Figs.~\ref{fig:fig3}\textit{A,B}), both the host lattice and the carrier sublattice are completely melted. Consequently, the trajectories exhibit homogeneous liquid-like motion at both short ($t = 1.5\tau_\alpha$) and long ($t = 40.0\tau_\alpha$) time windows, where $\tau_\alpha$ denotes the structural relaxation time corresponding to each temperature.
Upon lowering the temperature from the fully molten state into the sublattice melting regime (Figs.~\ref{fig:fig3}\textit{C,D}), the carrier trajectories begin to reflect the underlying crystalline environment imposed by the host lattice. While the short-time motion remains relatively localized, the long-time trajectories form a characteristic honeycomb-like pattern associated with diffusion through the interstitial pathways of the hexagonal host structure. This behavior indicates that the host lattice remains dynamically stable while simultaneously enabling long-range carrier transport through interconnected diffusion channels.
This apparent honeycomb pattern reflects carrier motion along the interstitial network defined by the host lattice, rather than hopping through host sites (see Supporting Information for a detailed visualization), demonstrating how the partially ordered host framework constrains and guides carrier diffusion.
\subsection*{Dynamical heterogeneity in solid ionics}
\begin{figure*}[t]
    \centering
    \includegraphics[width=\linewidth]{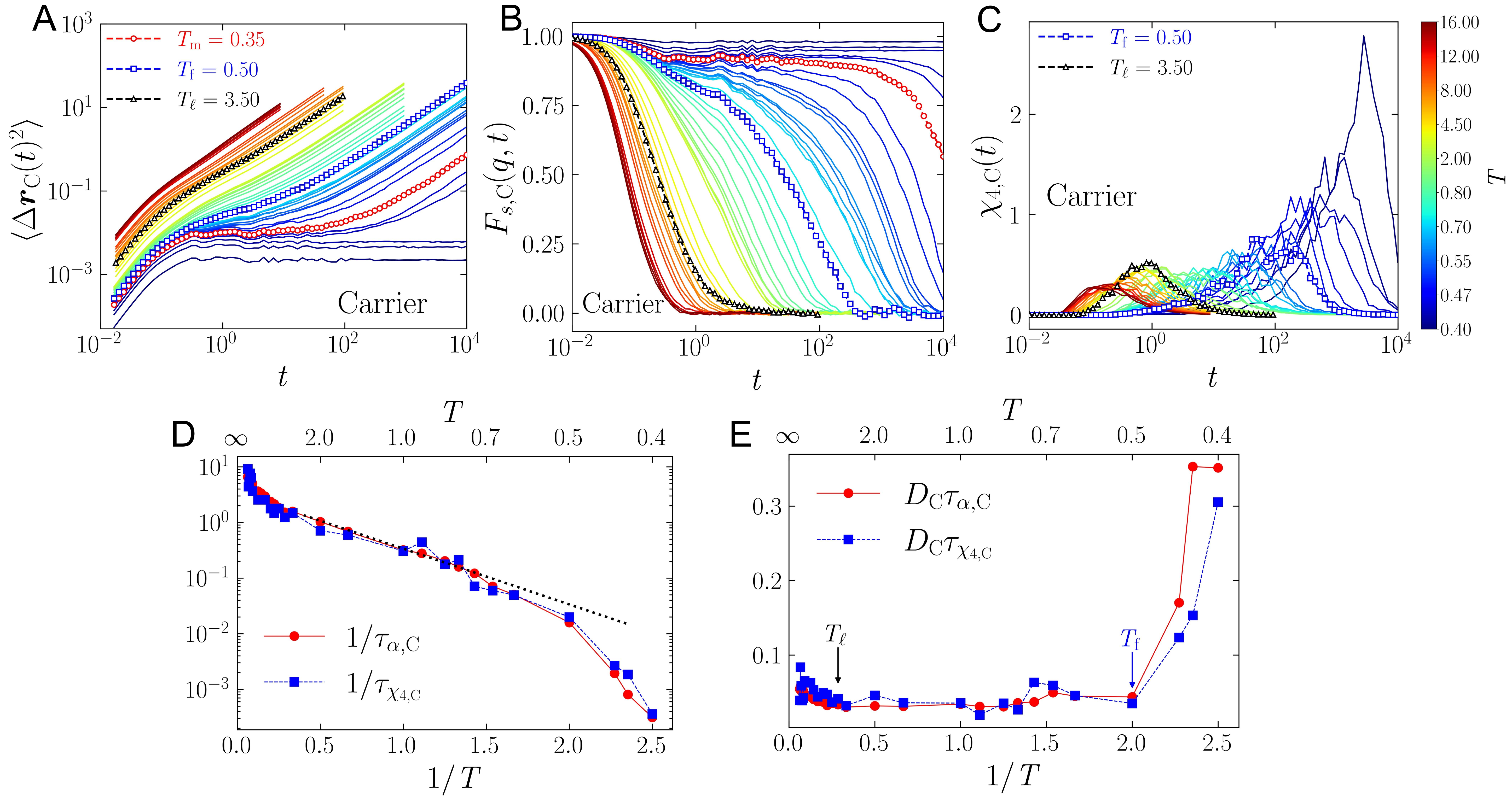}
    \caption{%
        \textbf{Microscopic dynamics and the extent of dynamical heterogeneity in 2D NAP model for $\phi = 0.85$:} (\textit{A-C}) Carrier dynamics are characterized by mean-squared displacement (MSD), self-intermediate scattering function $F_{\mathrm{s},\mathrm{C}}(q,t)$, and four-point susceptibility $\chi_{4,\mathrm{C}}(t)$ at different temperatures (indicated by color bar on the right). The characteristic temperatures $T_\mathrm{m} (\sim 0.35)$, $T_\mathrm{f} (\sim 0.50)$, and $T_\mathrm{\ell} (\sim 3.50)$-marking the onset of sublattice melting, the freezing of carrier motion, and the liquid-like regime, respectively-are indicated by red circles, blue squares, and black triangles connected with dotted lines. (\textit{D-E}) Inverse-temperature dependence of the Stokes-Einstein (SE) ratios $D_{\rm{C}}\tau_{\alpha, \mathrm{C}}$ and $D_{\rm{C}}\tau_{\chi_{4,\mathrm{C}}}$, highlighting pronounced SE violation at low temperatures (\textit{E}). A guideline in (\textit{D}) emphasizes the slope change near sublattice-melting transition.
    }
    \label{fig:fig4}
\end{figure*}

To elucidate the macroscopic dynamics in the system, we analysed several dynamic observables across different temperature regimes, as summarized in Fig.~\ref{fig:fig4}. The panels \textit{A-C} show the MSD of the carrier particles, the self-intermediate scattering function $F_{\mathrm{s}, \rm{C}}(q,t)$, and the four-point susceptibility $\chi_{4, \rm{C}}(t)$, respectively (see ``Materials and Methods''). For reference, three characteristic temperatures-$T_\mathrm{m} (\sim 0.35)$, $T_\mathrm{f} (\sim 0.50)$, and $T_\mathrm{\ell} (\sim 3.50)$-identified from the diffusivity behavior in Fig.~\ref{fig:fig2}, are marked in all panels. These represent, respectively, the onset of sublattice melting, the freezing of carrier motion, and the liquid-like regime.

The MSD data (Fig.~\ref{fig:fig4}\textit{A}) reveal that at low temperatures near $T_\mathrm{m}$, particles exhibit pronounced caged motion, which gradually weakens with increasing temperature and eventually dissolves above $T_\mathrm{f}$, leading to fully diffusive dynamics. Correspondingly, the self-intermediate scattering function $F_{\mathrm{s}, \rm{C}}(q,t)$ (Fig.~\ref{fig:fig4}\textit{B}), which measures the temporal correlation of particle positions at wavevector $q$, exhibits the emergence of a plateau around $T_\mathrm{f}$, signaling slow structural relaxation associated with carrier localization. At higher temperatures, $F_{\mathrm{s}, \rm{C}}(q,t)$ decays rapidly without any plateau, consistent with liquid-like mobility.

The four-point susceptibility for carriers $\chi_{4, \rm{C}}(t)$ (Fig.~\ref{fig:fig4}\textit{C}), which quantifies fluctuations in particle mobility and thus the extent of cooperative motion, captures the growth and suppression of correlated dynamics \cite{DauchotPRL2005, BerthierScience2005, LacevicJCP2003}.
At low temperatures, $\chi_{4,\rm C}(t)$ exhibits large and sharply peaked maxima at long times, indicating strong dynamical heterogeneity and persistent cooperative motion within the ordered carrier sublattice. As the temperature increases toward the sublattice melting regime near $T_\mathrm{f}$, the peak amplitude decreases rapidly and shifts to shorter times, reflecting the progressive loss of long-lived spatial correlations as the sublattice destabilizes. At still higher temperatures, only weak and broad peaks remain, consistent with increasingly homogeneous relaxation dynamics. Thus, the strong suppression of $\chi_{4,\rm C}(t)$ across the melting regime provides a microscopic signature of the crossover from heterogeneous cooperative transport to homogeneous diffusive motion.

To probe the connection between microscopic dynamics and transport, we examined the inverse relaxation times, $1/\tau_{\alpha, \mathrm{C}}$ and $1/\tau_{\chi_{4, \rm{C}}}$, as a function of inverse temperature $1/T$ (Fig.~\ref{fig:fig4}\textit{D}) (see ``Materials and Methods'' for the definition of these time scales). Both quantities exhibit nearly Arrhenius behavior at high temperatures, but deviate sharply near $T_\mathrm{f}$, indicating the onset of heterogeneous dynamics consistent with the MSD, $F_{\mathrm{s}, \rm{C}}(q,t)$, and $\chi_{4, \rm{C}}(t)$ results.
In the Arrhenius regime, relaxation is governed by an approximately fixed activation barrier $E_a$, such that the characteristic relaxation time or diffusivity follows a single Arrhenius slope. This implies that particles cross comparable energy barriers irrespective of the instantaneous configuration of their neighbors, a hallmark of weakly correlated dynamics typical of normal liquids or dilute systems.
However, near the sublattice melting transition, particle motion becomes increasingly cooperative and dynamically heterogeneous, with mobile and immobile regions coexisting in space and time. Structural relaxation therefore involves correlated rearrangements rather than isolated hopping events, leading to deviations from simple Arrhenius behavior in the relaxation dynamics. By contrast, purely local, single-particle hopping dynamics~\cite{KawasakiSciAdv2017} would preserve a nearly constant activation barrier and thus Arrhenius behavior even near the transition. Accordingly, the emergence of correlated motion naturally implies a breakdown of the Stokes--Einstein (SE) relation.

\begin{figure}[!t] 
    \centering
    \includegraphics[width=\linewidth]{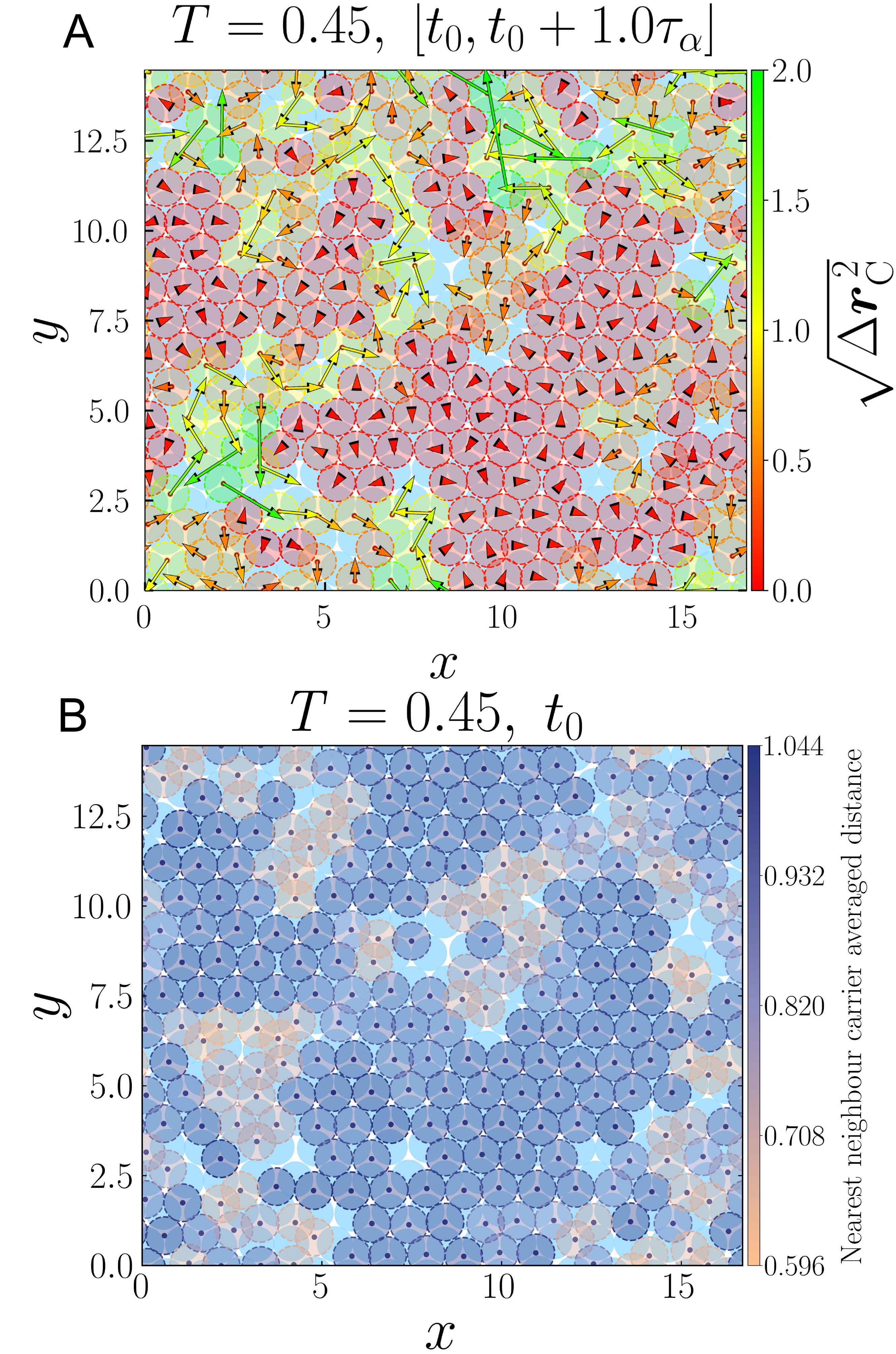}
    \caption{%
        \textbf{Interplay between dynamic and static heterogeneity near the sublattice-melting point in 2D NAP model for $\phi = 0.85$:} (\textit{A}) Spatial map of root mean-squared displacement of carriers over $t \in [t_0,\, t_0 + \tau_\alpha]$ at $T = 0.45$, showing coexisting liquid-like and immobile crystalline domains. Displacement vectors in (\textit{A}) further reveal string-like cooperative motion and hopping pathways within liquid-like regions, contrasted with localized, heterogeneous vibrations in crystallized domains. (\textit{B}) Static configuration at $t_0$, coloured by nearest carrier-carrier distance, revealing anharmonic regions that align with the dynamically mobile zones in (\textit{A}). Together, the two panels show that dynamical heterogeneity emerges from underlying static distortions of the carrier sublattice.
    }
    \label{fig:fig5}
\end{figure}

The corresponding Stokes-Einstein ratios $D_{\rm C}\tau_{\alpha,\mathrm{C}}$ and $D_{\rm C}\tau_{\chi_{4,\rm C}}$ (Fig.~\ref{fig:fig4}\textit{E}) remain nearly constant at high $T$, show a modest reduction near $T_\mathrm{\ell}$, and increase sharply below $T_\mathrm{f}$. This violation of the Stokes-Einstein relation is a hallmark of growing dynamical heterogeneity: structural relaxation slows markedly as correlated regions develop, whereas diffusion remains comparatively fast due to cooperative carrier motion. This violation of SE is consistent with nontrivial carrier-carrier correlations, which are commonly reflected in deviations of the Haven ratio from unity in heterogeneous transport regimes \cite{MurchSSI1982}. The resulting decoupling between the diffusion and relaxation time scales becomes pronounced once heterogeneous dynamics are long-lived \cite{KimJPSJ2010}, signaling the onset of sublattice melting. While these signatures establish the presence of dynamical heterogeneity \cite{KawasakiPRL2007, KawasakiSciAdv2017} from a macroscopic transport perspective, they do not resolve its microscopic origin. To directly probe the spatio-temporal structure of heterogeneous carrier motion near $T_\mathrm{f}$, we therefore examine particle-resolved displacements and local mobility patterns in Fig.~\ref{fig:fig5}.

To complement the dynamical observables discussed above, Figs.~\ref{fig:fig5}\textit{A,B} presents direct evidence of how structural and dynamical heterogeneity emerge near the sublattice-melting transition. 
Fig.~\ref{fig:fig5}\textit{A} maps the spatial distribution of carrier particle mobility through $\sqrt{\Delta \boldsymbol{r}_\mathrm{C}^{\,2}}$ over the interval $[t_0, t_0+\tau_\alpha]$ at $T=0.45$, further enhanced by displacement-vector arrows whose length and colour encode the magnitude of particle motion. These arrows reveal not only the coexistence of liquid-like and immobile domains, but also the presence of \emph{concerted hopping} events, manifested as string-like cooperative motion and correlated intermittent jumps within the mobile regions. Such collective displacements indicate that ion transport proceeds via cooperative rearrangements rather than independent single-particle hopping. 
This coexistence of mobile pathways and localized particles reflects pronounced dynamical heterogeneity at low temperature, which progressively diminishes with increasing temperature (see Supporting Information).
Notably, enhanced mobility is also observed near the edges of crystallized domains, reflecting pronounced local fluctuations of the carrier environment.
In contrast, the crystallized (red coloured) regions predominantly display short, irregular, and spatially heterogeneous displacement vectors, indicative of confined yet nonuniform vibrational dynamics. 
The corresponding static snapshot in Fig.~\ref{fig:fig5}\textit{B}, coloured by the local carrier-carrier nearest-neighbour averaged distance, exhibits a strikingly similar spatial pattern: particles in locally compressed, liquid-like environments (peach coloured) are spatially collocated with the dynamically mobile regions identified in Fig.~\ref{fig:fig5}\textit{A}, whereas particles in more expanded, ordered environments (navy blue coloured) coincide with dynamically frozen domains. Such a correspondence between structural softness and enhanced mobility closely parallels observations in glass-forming systems, where regions of local densification serve as loci of dynamical heterogeneity \cite{KawasakiPRL2007, TanakaNatMater2010}.
The emergence of such spatial heterogeneity is consistent with the breakdown of carrier hyperuniformity upon approaching $T_\mathrm{m}$, reflecting enhanced long-wavelength density fluctuations (see Supporting Information) \cite{TorquatoPhysRep2018}.
Notably, the peach-colored regions correspond to carrier particles transiently occupying interstitial positions within the host lattice. As previously proposed for superionic AgI, transient occupation of interstitial sites leads to strongly anharmonic vibrational motion rather than simple harmonic oscillations about lattice sites \cite{HoshinoSSC1977}. 
More broadly, recent studies in amorphous solids have shown that such emergent anharmonicity - manifested through low-frequency vibrational anomalies and instantaneous unstable modes-signals the proliferation of shallow, low-barrier sectors in the potential-energy landscape~\cite{YangPRB2022,KriuchevskyiPRE2022,ZacconeJCP2025}.
These insights may provide a natural microscopic framework for interpreting fast ionic transport in superionic conductors and its connection to the heterogeneous carrier dynamics observed in Fig.~\ref{fig:fig5}.
This directly motivates a quantitative analysis of lattice anharmonicity as the system approaches the sublattice-melting regime, which we pursue in the next section.
\subsection{Growth of anharmonicity near the sublattice melting point}
\begin{figure}[h]
    \centering
    \includegraphics[width=\linewidth]{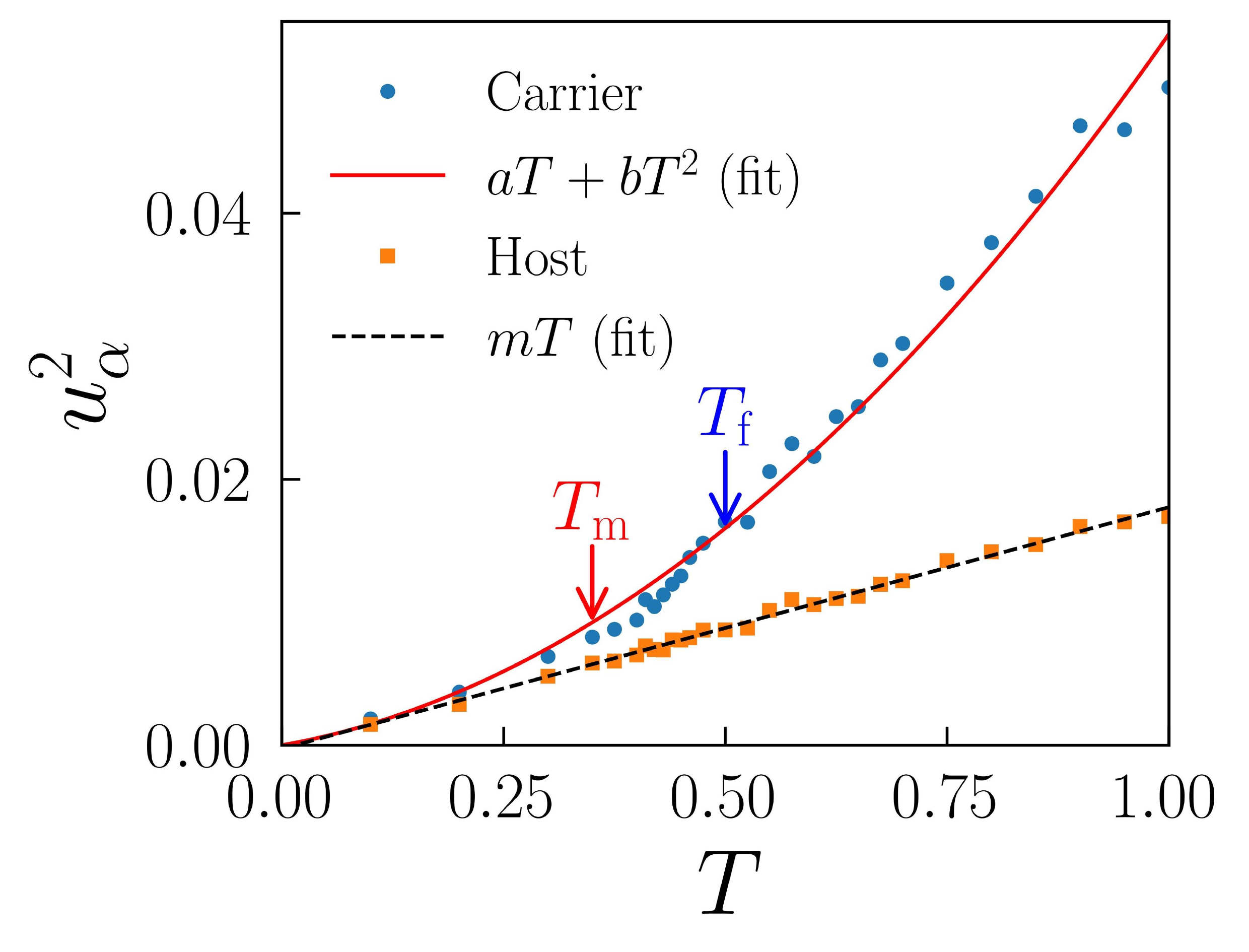}
    \caption{%
        \textbf{Anharmonic vibrational behavior from the Debye-Waller factor in 2D NAP model for $\phi = 0.85$:} Distinct anharmonic behavior is evident in the Debye-Waller factor: the host exhibits a linear dependence (black dotted line), indicating harmonic cage vibrations, whereas the carrier shows clear anharmonic effects (red solid line) with two distinct slopes, reflecting the increasing significance of nonlinear contributions near the onset of sublattice melting. 
    }
    \label{fig:fig6}
\end{figure}
\begin{figure}[!b]
    \centering
    \includegraphics[width=\linewidth]{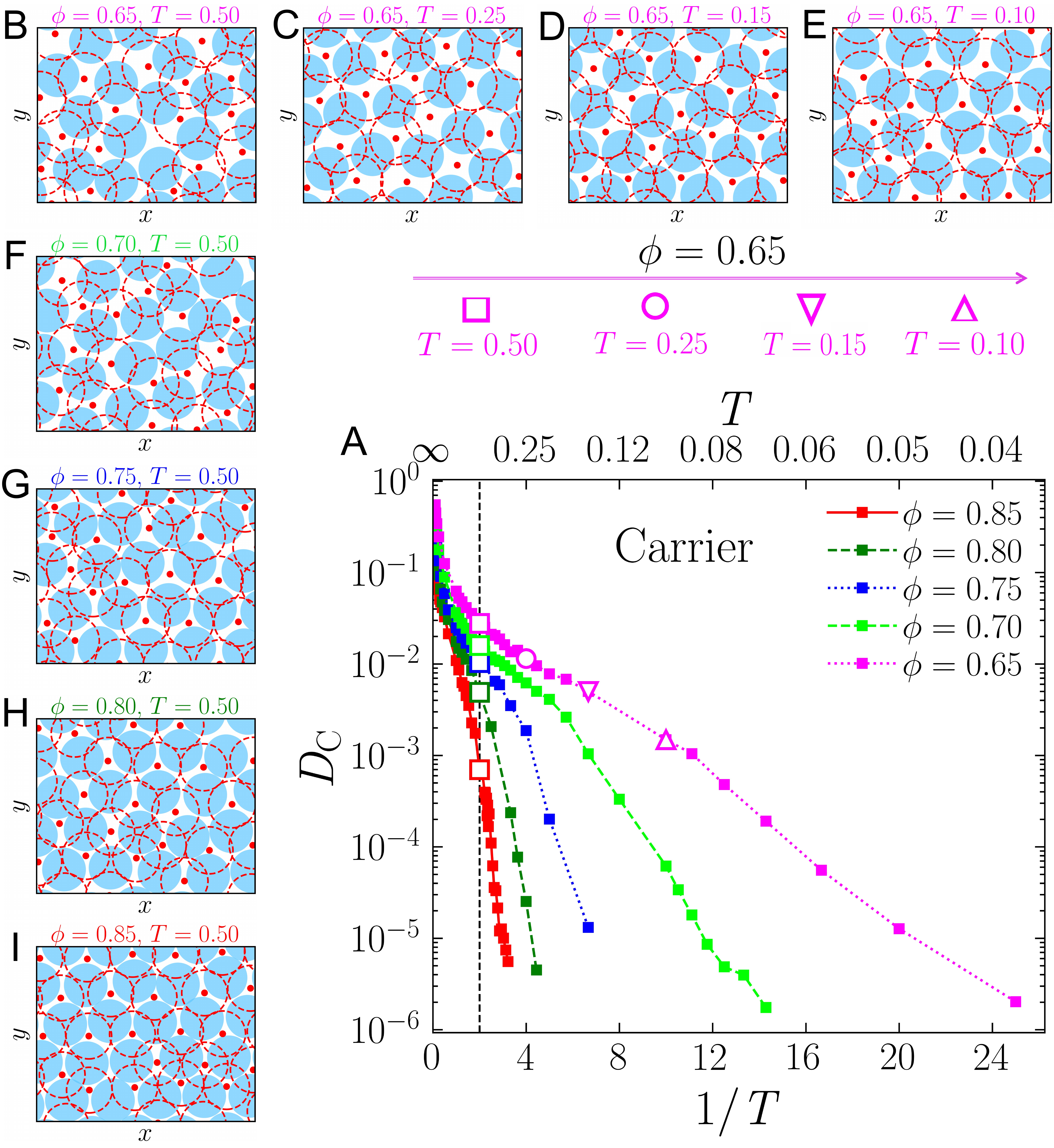}
    \caption{%
        \textbf{Density dependence of the diffusivity in 2D NAP model:} (\textit{A}) Temperature dependence of diffusivity of carriers ($D_{\mathrm{C}}$) as a function of inverse temperature ($1/T$) for different area packing fractions including $\phi=0.85$ (used for all previous results). The results demonstrate that tuning the packing fraction effectively controls the onset of sublattice melting. With decreasing packing fraction, the transition becomes smoother, and distinct slope changes are observed across the five packing cases, indicating different dynamical regimes. Panels (\textit{B--E}) correspond to $\phi = 0.65$ at decreasing temperatures ($T = 0.50, 0.25, 0.15, 0.10$), showing progressive localization of carriers upon cooling. Panels (\textit{B, F--I}) show configurations at $T = 0.50$ (follow the black dashed line in \textit{A}) for increasing densities ($\phi = 0.70-0.85$), demonstrating how higher packing suppresses anharmonic carrier motion and stabilizes the ordered sublattice. Colors are matched across diffusivity curves and corresponding configurations for each packing fraction.}
    \label{fig:fig7}
\end{figure}

To quantify the anharmonic vibrational state revealed above, we examine direct measures of lattice stability. In particular, we focus on the Debye-Waller factor, $u_\alpha^2$, defined as the mean-squared displacement of particles within their local cages, providing a measure of the vibrational amplitude that reflects local stiffness and the degree of anharmonicity in the lattice. In practice, $u_\alpha^2$ is extracted from the plateau value of the Lindemann index. A detailed analysis based on the Lindemann index is presented in the Supporting Information.

For the host particles, $u_{\rm H}^2$ increases linearly with temperature, $u_{\rm H}^2 \propto T$, indicating that their dynamics are governed by harmonic cage vibrations, as expected for an elastically stable lattice~\cite{KhrapakPRR2020}. 
We further find that the host lattice exhibits a density-dependent deviation from harmonic behavior: it remains approximately harmonic at high densities, whereas at lower densities $u_{\rm H}^2$ develops a clear nonlinear temperature dependence, indicating lattice softening and the onset of anharmonicity (see Supporting Information for more details).

In contrast, the carrier particles exhibit a pronounced nonlinear temperature dependence of the form $u_{\rm C}^2 = aT + bT^2$~\cite{KhrapakPRR2020}, with the quadratic contribution becoming increasingly significant as the system approaches the sublattice melting point. Within standard statistical-mechanical descriptions, such nonlinear behavior arises from anharmonic terms in the effective potential, signaling local softening and strongly anharmonic carrier fluctuations even while the host lattice remains elastically stable.

These results demonstrate that anharmonicity grows continuously upon approaching sublattice melting: the host retains predominantly harmonic, lattice-governed vibrations at high densities, whereas the carrier particles display pronounced nonlinear responses, reflecting the breakdown of harmonic confinement and the emergence of dynamically heterogeneous motion. 

\subsection{Controlling the sublattice melting}
We next examine how the onset of sublattice melting varies with density, characterized by the packing fraction in 2D (Fig.~\ref{fig:fig7}). The temperature dependence of the carrier diffusivity, $D_{\mathrm{C}}$, shows that decreasing packing fraction broadens the crossover between solid-like and fluid-like behavior.

This behavior indicates that the structural coherence of the host lattice weakens as density is reduced, consistent with a density-dependent lattice response in which lower densities exhibit earlier deviations from harmonic behavior, reflecting enhanced lattice softness that promotes more spatially extended and weakly correlated carrier dynamics. In contrast, higher densities maintain a more rigid lattice, leading to localized, heterogeneous motion and suppressed long-wavelength density fluctuations (see Supporting Information).

Consistent with this trend, both the host and carrier dynamics become increasingly anharmonic at lower densities (Figs.~\ref{fig:fig7}\textit{A--I}), reflecting the progressive softening of the lattice. This softening shifts the onset of sublattice melting to lower temperatures, facilitates carrier diffusion, and reduces the degree of dynamical heterogeneity.

\subsection{Generality of dynamical features: $\alpha$-AgI}
\begin{figure}[h]
    \centering
    \includegraphics[width=\linewidth]{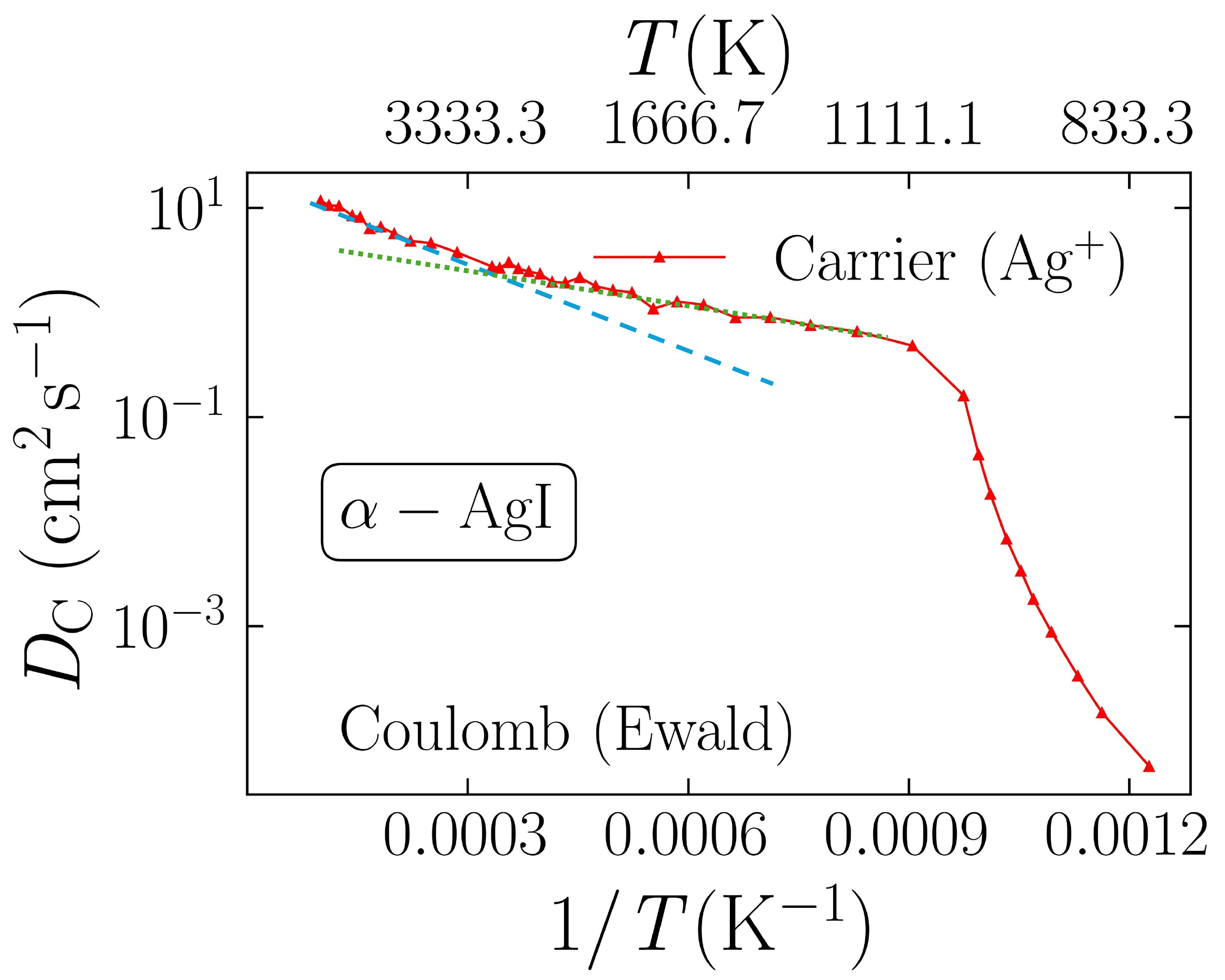}
    \caption{%
        \textbf{Diffusivity of 3D $\alpha$-AgI (FUH) model as a function of inverse temperature:} Results from molecular dynamics simulations with Coulomb interactions treated via Ewald summation. Changes in slope indicate distinct dynamical regimes within the superionic phase, consistent with sublattice melting and the onset of diffusive motion, analogous to the behavior shown in Fig.~\ref{fig:fig2}\textit{A} and Fig.~\ref{fig:fig7}\textit{A}.
    }
    \label{fig:fig8}
\end{figure}
To assess the generality of the dynamical features observed in our minimal 2D~NAP model, we performed molecular dynamics simulations of 3D $\alpha$-AgI using a more realistic interaction potential (FUH model) that combines short-range repulsion with long-range Coulomb interactions treated via the Ewald summation method~\cite{FukumotoJPSJ1982} (see ``Materials and Methods''). 
The resulting temperature dependence of the diffusivity (Fig.~\ref{fig:fig8}) exhibits qualitative features closely resembling those of our 2D~NAP model (Fig.~\ref{fig:fig2}\textit{A} and Fig.~\ref{fig:fig7}\textit{A}). Importantly, although the experimental AgI system exhibits a first-order structural transition, the dynamical behavior within the superionic phase remains consistent with that identified in our minimal model. In particular, distinct transport regimes associated with sublattice melting and fully developed diffusive motion, together with the crossover in diffusivity reflected by changes in slope in the Arrhenius representation, are observed independently of whether the underlying transition is continuous or discontinuous.
This indicates that the microscopic transport mechanism-namely, the emergence of cooperative carrier motion driven by anharmonic lattice fluctuations-is not tied to the nature of the phase transition itself. Rather, the transition sharpness reflects thermodynamic aspects of lattice reconstruction, whereas the transport behavior is governed by dynamical processes within the superionic phase.

Because the 2D~NAP model does not incorporate lattice reconstruction, it is not intended to reproduce the discontinuous structural transition itself, but rather the universal dynamical features of superionic transport.

\section*{Summary}
We have investigated a minimal two-dimensional binary mixture model of a superionic conductor to disentangle three intertwined ingredients of fast ion transport-\emph{sublattice melting}, \emph{dynamical heterogeneity}, and \emph{anharmonicity}-and to clarify how the density governs their emergence.  
The 2D~NAP model captures selective sublattice melting within a chemically neutral, coarse-grained framework.  
Host (H) particles interact via short-range steric repulsion, forming a rigid lattice, whereas carrier (C) particles interact through much softer, long-range Wigner-type forces that promote collective delocalization.  
This asymmetry in interaction range naturally produces distinct melting temperatures for the two sublattices, enabling a selective carrier-melting phase where the host remains ordered while the carrier sublattice becomes fluid-like (Fig.~\ref{fig:fig1}).  
Brownian dynamics was used for equilibration, followed by long microcanonical molecular dynamics simulations to capture intrinsic dynamics (see ``Materials and Methods'').

Snapshots and radial distribution functions reveal clear \emph{selective disordering}: the carrier sublattice loses positional order while the host remains crystalline (Figs.~\ref{fig:fig1}, \ref{fig:fig2}).  
Diffusivity-temperature plots identify four regimes (Fig.~\ref{fig:fig2}\textit{A}): (I) crystalline, (II) onset of sublattice melting, (III) a broad \emph{sublattice-melting} region with mobile carriers in an ordered host, and (IV) full melting.  
Diffusivity of carriers evolves smoothly across regime II, whereas the host shows a sharp discontinuity, consistent with a \emph{first-order-like} transition (Fig.~\ref{fig:fig2}\textit{A}). Carrier trajectories (Fig.~\ref{fig:fig3}) visualize these regimes.  
At high $T$, motion is homogeneous and isotropic. Within regime II, carriers move cooperatively along host-lattice-aligned pathways forming transient honeycomb-like networks, while at lower $T$ they exhibit heterogeneous motion, alternating between localized and collective migration. These patterns confirm that sublattice melting arises from correlated, anharmonic motion rather than uncorrelated hopping.

Time-correlation analyses quantify these dynamics (Fig.~\ref{fig:fig4}). The mean-squared displacement and self-intermediate scattering function reveal transient caging and stretched-exponential relaxation, marking cooperative slow dynamics. The four-point dynamic susceptibility $\chi_{4,\mathrm{C}}(t)$ peaks near the freezing side of regime~II, indicating growing spatial correlations, and collapses in the liquid phase, signifying a crossover from heterogeneous to homogeneous relaxation.  
The Stokes-Einstein ratios for carriers deviate strongly near the transition, confirming the emergence of cooperative transport beyond single-particle diffusion. 

Spatially resolved mobility maps reveal pronounced dynamical heterogeneity near the sublattice-melting transition (Fig.~\ref{fig:fig5}\textit{A}). Carriers segregate into coexisting mobile and immobile domains, with string-like cooperative displacements and correlated hopping events indicating collective, rather than single-particle, transport. Enhanced mobility preferentially appears near the boundaries of crystallized regions, where particles transiently occupy interstitial positions.
The corresponding static snapshots (Fig.~\ref{fig:fig5}\textit{B}) demonstrate a strong correspondence between local structural softness and mobility: dynamically active regions coincide with locally compressed, liquid-like environments, while dynamically frozen regions remain structurally ordered. This spatial correlation parallels the behavior observed in glass-forming systems and implicates \emph{anharmonic lattice distortions} as the \emph{microscopic origin of heterogeneous carrier dynamics} near sublattice melting.

The \emph{Debye-Waller factor} links these dynamical changes to local vibrations (Fig.~\ref{fig:fig6}). At high density, the host particles exhibit nearly harmonic behavior, whereas the carrier particles display increasingly nonlinear vibrational dynamics near the sublattice melting regime, signaling enhanced anharmonicity. Reducing the packing fraction broadens regime~III, softens the host lattice, and induces increasingly nonlinear fluctuations in both subsystems (Fig.~\ref{fig:fig7}), demonstrating that density controls the cooperativity of sublattice melting by reshaping the local energy landscape.
Simulations of 3D $\alpha$-AgI with Coulomb interactions (Fig.~\ref{fig:fig8}) reproduce qualitatively similar transport regimes, supporting the broader relevance of the mechanisms identified in the minimal model.

These insights suggest that \emph{sublattice melting can be engineered via density control}, offering design principles for robust, high-conductivity solid electrolytes. The roles of dimensionality and carrier size in shaping transport behavior are examined in detail in the Supporting Information.

\section*{Discussions}
While the present 2D~NAP model captures the essential physics of sublattice melting, several directions remain for future exploration. 
Although the density-controlled sharp change in transport shows features often associated with phase transitions, its precise thermodynamic classification is subtle. We therefore use sublattice melting to denote the underlying mechanism-selective carrier delocalization driven by enhanced anharmonicity and spatially heterogeneous, string-like motion-rather than a strict phase-transition taxonomy.
Importantly, this dynamical heterogeneity directly facilitates transport by generating interconnected high-mobility regions that enable cooperative carrier motion, shifting transport control from average energy barriers to the connectivity of these pathways.
Within this framework, our results are consistent with prior studies reporting barrier reduction due to collective ion motion \cite{HeNatCommun2017}. Here, we show that this reduction arises from an entropic contribution, whereby collective dynamics enhance the migration entropy and thereby lower the effective activation barrier governing transport. A detailed decomposition of energetic and entropic contributions is provided in the Supporting Information.
While string-like cooperative dynamics and non-Arrhenius ionic transport have been reported in more realistic models~\cite{WangACSAEM2021}, we show within our minimal NAP model that such behaviors arise from a unified mechanism in which anharmonic lattice fluctuations induce spatially heterogeneous dynamics near the onset of sublattice melting.
In the high-temperature superionic state, carrier motion is constrained by bottlenecks imposed by the host lattice but remains dynamically homogeneous and liquidlike. Near the onset of sublattice melting, however, the dynamics become strongly heterogeneous, with transient coexistence of temporarily localized and liquidlike carrier regions, reminiscent of cooperative motion in glass-forming systems. This heterogeneity is progressively suppressed as the system evolves into the fully developed superionic regime.
Although the model is formulated in two dimensions for simplicity and visualization, the observed dynamical behavior is not restricted to dimensionality or strict long-range translational order. Our chemistry-neutral minimal NAP model isolates the essential dynamical mechanism while omitting explicit long-range electrostatics and polarizability. Incorporating these effects is a natural extension and is expected to shift quantitative thresholds without altering the qualitative mechanism. Indeed, corresponding 3D simulations of $\alpha$-AgI reproduce qualitatively similar diffusivity crossovers and transport regimes within the superionic phase, supporting the generality of the underlying dynamical behavior. 
To further clarify the microscopic origin of the transport regimes, we analyze the carrier dynamics of the corresponding 3D NAP system in the Supporting Information. These simulations reproduce the same qualitative sequence of transport crossovers observed in two dimensions and further demonstrate that carrier size strongly influences the dominant high-temperature transport mechanism. In particular, comparatively larger carriers exhibit an additional activated regime associated with the onset of host-lattice melting, indicating a separation between carrier-controlled and host-controlled transport processes. Together, these results support the generality of the underlying dynamical picture across dimensionality and interaction regimes.
To rigorously validate the microscopic mechanism proposed here, it is crucial to connect this coarse-grained framework with first-principles descriptions.

Experimentally, our findings offer clear guidelines for materials design. In real systems, the effective density-or equivalently, the lattice volume or chemical pressure-can be tuned through compositional substitution, strain, or external pressure; these controls correspond to effective lattice expansion and defect engineering that modulate free volume and local lattice stiffness.  
We predict that reducing the effective density broadens the sublattice-melting regime and enhances carrier anharmonicity, implying that low-density or expanded-lattice phases may exhibit sublattice melting at comparatively lower temperatures.  
Neutron and X-ray scattering measurements provide access to carrier diffusion, intermediate scattering functions, and Debye-Waller factors, enabling characterization of relaxation dynamics, cooperative motion, and the predicted crossover from harmonic to anharmonic behavior.
Raman and quasielastic neutron spectroscopy can further probe phonon softening and correlated ionic motion for quantitative comparison with simulations.
Signatures of dynamical heterogeneity may also be inferred from deviations from the Stokes-Einstein relation, obtained by combining diffusion measurements with spectroscopic relaxation times.
Importantly, this framework provides a direct correspondence between microscopic dynamical processes-such as anharmonic fluctuations and cooperative carrier motion-and experimentally measurable transport properties, including ionic diffusivity and non-Arrhenius behavior, thereby providing a physically grounded interpretation of superionic conduction across a broad class of materials.

To further assess the quantitative relevance of the minimal NAP model beyond qualitative phenomenology, we compared the carrier diffusivity obtained at packing fraction $\phi = 0.70$ with experimental diffusion data for $\beta$-alumina in the Arrhenius regime associated with sublattice melting~\cite{KamishimaJPSJ2010} (see Supporting Information for details). 
Although $\beta$-alumina contains distinct crystallographic diffusion sites (e.g., BR and a-BR) and associated energetic heterogeneity, our NAP model assumes equivalent sites and therefore isolates the underlying collective transport mechanism rather than reproducing material-specific microscopic details. 
By collapsing the Arrhenius regimes of the simulated and experimental diffusivity curves, we obtain an effective activation scale $\epsilon/k_\mathrm{B} = 1264.3~\mathrm{K}$ and a transport prefactor $a^2/\tau_0 = 2.05 \times 10^{-5}~\mathrm{cm}^2\,\mathrm{s}^{-1}$. These results further support the robustness of the underlying transport mechanism captured by the NAP model. 
Details of the fitting procedure, scaling, and physical interpretation of these parameters are provided in the Supporting Information. 

In conclusion, this study provides a quantitative and conceptually transparent framework that bridges phenomenological modeling with an atomistic understanding of sublattice melting. 
Unlike conventional mean-field or single-ion hopping descriptions-which treat ionic transport as independent motion in a static potential-our results reveal that fast ion conduction arises from \emph{collective, anharmonic, and density-dependent} dynamics constrained by the crystalline host, offering a microscopic foundation for interpreting nonlinear transport behaviors observed in superionic conductors.  
Combining first-principles simulations with controlled experiments will be essential to verify the universality of these mechanisms and translate them into design principles for next-generation solid electrolytes.

\section{Materials and Methods}
Molecular dynamics simulation codes were developed in C++ (2D and 3D NAP model) and LAMMPS (3D $\alpha$-AgI model) \cite{ThompsonCPC2022}, and post-processing, analysis, and figure generation were carried out using Python scripts with Matplotlib \cite{HunterMatplotlib2007} and OVITO \cite{StukowskiOVITO2010}.
\subsection{SIMULATIONS}
\subsection{Two-dimensional non-additive potential (NAP) model: interactions and simulation protocol} To reproduce the essential physics of sublattice melting within a minimal and chemically neutral framework, we impose distinct interaction types on the two species. The host particles interact through short-range steric repulsion, forming a mechanically rigid crystalline framework, whereas the carrier particles experience a much softer, effectively long-range Wigner-type interaction that promotes collective delocalization within the host lattice. This asymmetry in interaction range and stiffness naturally yields widely separated melting temperatures for the two sublattices, enabling the emergence of a selective carrier-melting phase in which the carrier sublattice becomes fluidlike while the host lattice remains ordered. We refer to this minimal, chemically neutral framework as the non-additive potential (NAP) model.

To place our approach in context, prior studies of superionic conductors have shown that ion transport and sublattice-selective melting can emerge from relatively simple interaction schemes. In tunnel-type solids, a balance of Coulombic, polarization, and short-range repulsive forces governs the off-axis migration pathways and activation barriers of mobile ions, producing size-selective mobility \cite{FlygareHugginsJPCS1973, FukumotoJPSJ1982}. 
More broadly, these studies demonstrate that coarse-grained molecular dynamics with simplified but physically motivated interactions can reproduce essential features of ion-conducting phases. Our NAP model follows this spirit but adopts an even more generalized, chemically neutral formulation, retaining the core interplay between host confinement and carrier delocalization while avoiding system-specific assumptions.

\begin{table}[b]
\caption{\label{tab:params}
Model parameters used in the simulations.
}
\begin{ruledtabular}
\begin{tabular}{lcccccccc}
$N_{\mathrm{C}}{:}N_{\mathrm{H}}$ &
$m_{\mathrm{C}}$ &
$m_{\mathrm{H}}$ &
$\epsilon_{\mathrm{CC}}$ &
$\epsilon_{\mathrm{HH}}$ &
$\epsilon_{\mathrm{CH}}$ &
$\sigma_{\mathrm{C}}$ &
$\sigma_{\mathrm{H}}$ &
$\sigma_{\mathrm{CH}}$ \\
\colrule
1{:}1 & 1.0 & 1.0 & 0.001 & 1.0 & 1.0 & 0.154 & 1.0 & 0.577 \\
\end{tabular}
\end{ruledtabular}
\end{table}

To this end, we consider a two-dimensional binary mixture comprising a total of $N = 512$ particles, with equal numbers of host ($N_\mathrm{H} = 256$) and carrier ($N_\mathrm{C} = 256$) species. The moderate system size is intentionally chosen to suppress long-wavelength fluctuations associated with the Mermin-Wagner theorem in two dimensions, while preserving the intrinsic transport mechanisms of interest~\cite{ShibaPRL2016, ShibaJPCM2018, ShibaPRL2019}. As shown in Fig.~\ref{fig:fig1}\textit{B}, the initial configuration of the binary mixture is constructed in a hexagonal packing arrangement, where the host particles define the underlying lattice and the carriers are distributed within the interstitial sites. The interparticle interactions are modeled using the Weeks-Chandler-Andersen (WCA) potential, a truncated and shifted form of the Lennard-Jones potential, as detailed in the following.

For a pair of particles $i$ and $j$ separated by a distance $r_{ij} = |\mathbf{r}_i - \mathbf{r}_j|$, the interaction potential is given by 
$U = \sum_{\langle i,j \rangle} U_{ij}(r_{ij})$ where, 

\begin{equation}
\begin{aligned}
U_{ij}(r_{ij}) &=
\begin{cases}
4 \epsilon_{ij} \left[
\left( \dfrac{\sigma_{ij}}{r_{ij}} \right)^{12}
- \left( \dfrac{\sigma_{ij}}{r_{ij}} \right)^6
\right] + \epsilon_{ij}, & r_{ij} \le r_c, \\[8pt]
0, & r_{ij} > r_c.
\end{cases}
\end{aligned}
\label{eq:WCA}
\end{equation}

where the notation $\langle i,j\rangle$ denotes a sum over all distinct particle pairs ($i\neq j$), with each pair counted only once. $\sigma_{ij}$ is the effective particle diameter for the interacting pair and $\epsilon_{ij}$ sets the energy scale. The cutoff distance $r_{c}$ = $2^{1/6}\sigma_{ij}$ ensures that the potential is purely repulsive. In our simulation, we use $\sigma_{\mathrm{H}}$, $\epsilon_{\mathrm{HH}}$, $m_{\mathrm{H}}$, and $\sqrt{m_{\mathrm{H}}\sigma_{\mathrm{H}}^2/\epsilon_{\mathrm{HH}}}$ as units of length, energy, mass, and time, respectively. 
Temperature is expressed in reduced units as $T^* = k_{\mathrm{B}}T/\epsilon_{\mathrm{HH}}$, where $\epsilon_{\mathrm{HH}}$ sets the characteristic energy scale of the system. A correspondence with real temperatures is established by matching the Arrhenius behavior of the simulated diffusivity to experimental data (see below and Supporting Information).

The interaction parameters are symmetric, i.e., $\sigma_{ij}=\sigma_{ji}$ and $\epsilon_{ij}=\epsilon_{ji}$, and the corresponding values of $(\sigma_{ij}, \epsilon_{ij})$ for all particle pair types are summarized in Table~\ref{tab:params} (see Fig.~\ref{fig:fig1}\textit{A} for detailed visualization). The host-host ($\mathrm{HH}$) and carrier-host ($\mathrm{CH}$) interactions are defined with a common energy scale $\epsilon_{ij}=1.0$, whereas the carrier-carrier ($\mathrm{CC}$) interaction is deliberately chosen to be much weaker, $\epsilon_{\mathrm{CC}}=10^{-3}$. 
This parametrization suppresses direct mutual exclusion between carriers and promotes collective dynamics within the confining host lattice, consistent with earlier coarse-grained descriptions of superionic transport based on long-range correlations rather than explicit short-range repulsion \cite{FukumotoJPSJ1982, LinPNAS2023}.
In contrast to those works, however, we do not introduce explicit charge-specific Coulomb interactions. Instead, \emph{we employ a non-additive, long-range interaction in a minimal form}, designed to capture the essential collective effects of carrier motion while remaining chemically neutral. 

In this minimal framework, explicit electrostatic and polarization effects are neglected to isolate the roles of interaction asymmetry and lattice softness in sublattice melting and carrier dynamics. While such effects may alter quantitative scales (e.g., transition temperatures and transport coefficients), the qualitative features remain robust, as confirmed by simulations with more realistic interaction potentials (see below).

Accordingly, our two-dimensional NAP model captures key features of superionic behavior, including selective sublattice melting and strongly heterogeneous carrier dynamics, indicating that these phenomena arise primarily from collective effects governed by lattice softness and long-range correlations among mobile carriers.


\begin{table}[t]
\caption{\label{tab:packing}
Variation of simulation box parameters with area packing fraction $\phi$.
}
\begin{ruledtabular}
\begin{tabular}{ccccc}
$\phi$ &
$a_{\mathrm{L\text{-}space}}$ &
$\sigma_{\mathrm{CC}}$ &
$L_x$ &
$L_y$ \\
\colrule
0.85 & 1.0440 & 1.0440 & 16.7040 & 14.4661 \\
0.80 & 1.0773 & 1.0773 & 17.2362 & 14.9270 \\
0.75 & 1.1126 & 1.1126 & 17.8016 & 15.4166 \\
0.70 & 1.1516 & 1.1516 & 18.4256 & 15.9570 \\
0.65 & 1.1951 & 1.1951 & 19.1216 & 16.5598 \\
\end{tabular}
\end{ruledtabular}
\end{table}

The lattice spacing $a_{\mathrm{L\text{-}space}}$, effective carrier size $\sigma_{\mathrm{CC}}$, and the corresponding box dimensions $(L_x, L_y)$ for the two-dimensional system at different area packing fractions $\phi$ are summarized in Table~\ref{tab:packing}.

We employed both underdamped Brownian dynamics (BD) and Newtonian molecular dynamics (MD) to evolve the binary mixture. 
During the initial equilibration stage, the system was evolved under underdamped Langevin dynamics, $m_i \frac{{\rm d} \mathbf{v}_i}{{\rm d}t} = - \zeta \mathbf{v}_i - \frac{\partial U}{\partial \mathbf{r}_i} + \boldsymbol{\eta}_i(t)$, where $\zeta$ is the damping coefficient and $\boldsymbol{\eta}_i(t)$ is a Gaussian white noise term with zero mean: $\langle \boldsymbol{\eta}_i(t) \rangle = \mathbf{0}$  and the variance $\langle \boldsymbol{\eta}_{i}(t)\otimes\boldsymbol{\eta}_{j}(t') \rangle = 2 k_\mathrm{B}T \zeta \, 
\delta_{ij}\delta(t-t')\mathbf{1}$ due to fluctuation-dissipation theorem and $\mathbf{1}$ denotes the unit matrix. BD simulations were propagated for $2\times 10^{6}$ steps with timestep $\Delta t_\mathrm{BD} = 0.001$, corresponding to a total equilibration time $t_\mathrm{BD} = 2\times 10^3$. After equilibration, the system was switched to microcanonical molecular dynamics (NVE ensemble) to follow the Hamiltonian evolution. MD simulations were performed for $1 \times 10^{8}$ steps with the time step $\Delta t_\mathrm{MD} = 0.0001$, corresponding to a total production run length $t_\mathrm{MD} = 1\times10^{4}$.

In both stages, the system of $N$ particles was propagated for sufficiently long times to ensure equilibration and statistical averaging. Brownian dynamics ensured relaxation into equilibrium configurations, whereas subsequent NVE molecular dynamics enabled the study of intrinsic dynamical processes without thermostatting.

\subsection{Radial distribution function}
The pair distribution function (PDF) of particles of type $\alpha \in \{{\rm C, H}\}$ (carrier or host) is defined \cite{HansenMcDonald2013} as  
$g_{\alpha\alpha}(\mathbf{r}) =
\frac{1}{\rho_{\alpha} N_\alpha}
\left\langle
\sum_{j \neq k,\; j,k \in \alpha}
\delta\!\left( \mathbf{r} - \mathbf{r}_j + \mathbf{r}_k \right)
\right\rangle,$
where $\rho_{\alpha} = N_{\alpha}/(L_x L_y)$ is the number density of particle of species $\alpha$ and $N_\alpha$ is the number of particles of species $\alpha$.  
The radial distribution function (RDF), $g_{\alpha\alpha}(r)$, was obtained by averaging the PDF over all angular directions.  
This function quantifies the local structural correlations within each particle species and provides a measure of the characteristic interparticle spacing and short/long-range order.

\subsection{Mean squared displacement}
The mean-squared displacement (MSD) of particles of type $\alpha \in \{{\rm C, H}\}$ (carrier or host) was computed as \cite{KobPRE1995}  
$
\langle \Delta \mathbf{r}_\alpha(t)^2 \rangle
= \frac{1}{N_\alpha}
\sum_{j \in \alpha}
\left\langle
\big[ \mathbf{r}_j(t+t_0) - \mathbf{r}_j(t_0) \big]^2
\right\rangle .
$
At long times, the MSD exhibits diffusive behavior, from which the self-diffusion coefficient $D_\alpha$ is determined as  
$
\langle \Delta \mathbf{r}_\alpha(t)^2 \rangle \sim 4 D_\alpha t .
$
This analysis was conducted separately for the carrier and host species, allowing a direct comparison of their respective mobilities and of the temperature dependence of their transport dynamics.

\subsection{Self-intermediate scattering function}
The self-intermediate scattering function of particles of type $\alpha \in \{{\rm C, H}\}$ (carrier or host) was calculated as \cite{KobPRL1994, HansenMcDonald2013}  
$
F_{\mathrm{s}, \mathrm{\alpha}}(q,t)
= \frac{1}{N_{\mathrm{\alpha}}}
\sum_{j \in \mathrm{\alpha}}
\left\langle
\exp \!\left[
i \mathbf{q} \cdot
\big( \mathbf{r}_j(t+t_0) - \mathbf{r}_j(t_0) \big)
\right]
\right\rangle,
$
where $\mathbf{q}$ is the wave vector, whose magnitude was chosen to correspond to the characteristic length scale of the host particles, i.e., near the first peak of the static structure factor.  
Relaxation dynamics was characterized by monitoring the temporal decay of $F_{\mathrm{s}, \mathrm{\alpha}}(q,t)$. In the following, we restrict the analysis to the carrier species. 
The structural relaxation time for carriers $\tau_{\alpha, \mathrm{C}}$ was determined by fitting $F_{\mathrm{s}, \mathrm{C}}(q,t)$ to the Kohlrausch-Williams-Watts (KWW) function,  
$F_{\mathrm{s}, \mathrm{C}}(q,t) \sim A \exp \{-[t / \tau_{\alpha, \mathrm{C}}]^{\beta} \}$,
where $A$ and $\beta$ are the fitting parameters that represent the amplitude and stretching exponent, respectively.  
This analysis was performed to directly probe the microscopic relaxation and temperature-dependent dynamics of the carriers.

\subsection{Four-point dynamic susceptibility}
To quantify dynamical heterogeneity consistent with our analysis of $F_{\mathrm{s},\alpha}(q,t)$, we evaluated the four-point dynamic susceptibility \cite{LacevicJCP2003} for particles of type $\alpha \in \{{\rm C, H}\}$ (carrier or host) using the self-overlap order parameter.
The single-particle overlap is defined as
$w^{(\alpha)}_j(t_0{+}t,t_0)
\equiv
\Theta\!\Big(a - \big|\mathbf{r}_j(t_0{+}t)-\mathbf{r}_j(t_0)\big|\Big)$, 
where $\Theta(\cdot)$ is the Heaviside step function and $a$ is a microscopic cutoff that sets the cage scale.
Unless otherwise noted, we set $a = a_{\text{L-space}}/(2\sqrt{3})$, a geometric threshold corresponding to half the nearest-neighbor spacing in the host lattice, ensuring that the overlap function probes cage-breaking displacements on the host lattice length scale.

The species-resolved (self) overlap is
$Q_\alpha(t) = \sum_{j\in \alpha} w^{(\alpha)}_j(t_0{+}t,t_0)$,
with averages taken over $t_0$ (multiple time origins) and independent trajectories.
The four-point dynamic susceptibility is then obtained from the variance of $Q_\alpha(t)$ as
$
\chi_{4,\alpha}(t) = \frac{1}{N_\alpha}
\left( \langle Q_\alpha(t)^2 \rangle - \langle Q_\alpha(t) \rangle^2 \right)
$.
In reporting $\chi_{4,\alpha}(t)$, we highlight a characteristic time: the peak time $\tau_{\chi_4}$ that maximizes $\chi_{4,\alpha}(t)$.
All quantities were computed separately for carrier particles only to compare the species-resolved growth of dynamic correlations and their temperature dependence.

\subsection{Lindemann index and Debye-Waller factor}
To quantify species-resolved local relative fluctuations in particle positions, we computed the two-dimensional Lindemann index \cite{BedanovPLA1985, FanActaMater2020, GilvarryPR1956a, GilvarryPR1956b, KhrapakPRR2020, LuoJCP2005, VopsonSSC2020, WolfJGR1984} $\gamma_{{\rm L},\alpha}(t)$ for each component $\alpha \in \{{\rm C, H}\}$ (carrier or host).  
Following the definition of the relative displacement between particles $j$ and $k$,
$\mathbf{u}_{jk}(t)
= \big[\mathbf{r}_j(t) - \mathbf{r}_k(t)\big]
  - \big[\mathbf{r}_j(0) - \mathbf{r}_k(0)\big]$,
the Lindemann index for each species is defined as
$
\gamma_{{\rm L},\alpha}(t)
= \frac{1}{N_\alpha}
  \sum_{j \in \alpha}
  \frac{1}{n_j}
  \sum_{k \in \alpha,~{\rm n.n.}(j)}
    \big\langle
      \big|\mathbf{u}_{jk}(t)\big|^2
    \big\rangle/a_{\rm L-space}^2
  .
$
Here, ${\rm n.n.}(j)$ denotes the set of first-nearest-neighbor host particles surrounding particle $j$.  
The integer $n_j$ is the coordination number of particle $j$, determined from the first minimum of the host-host radial distribution function.  
The Lindemann index $\gamma_{{\rm L},\alpha}(t)$ thus measures the relative amplitude of the local vibrational motion of $\alpha$ particles with respect to their nearest-neighbor distances on the host-particle length scale.  
This function was computed independently for carrier and host particles, allowing direct comparison of their local structural stability and the temperature dependence of their relative fluctuations.

The Debye--Waller factor quantifies the amplitude of vibrational motion in the caged regime and is closely related to the plateau value of the Lindemann index.
For each species $\alpha \in \{{\rm C},{\rm H}\}$ (carrier or host), we define
$
u_\alpha^2
=  \gamma_{L,\alpha}(t_{\rm p}),
$
where $\gamma_{L,\alpha}(t_{\rm p})$ denotes the value taken at the characteristic turning time $t_{\rm p}$ between the ballistic and diffusive regimes.
This time marks the cage-confinement timescale at which both the mean-squared displacement and $\gamma_{L,\alpha}(t)$ exhibit a transient plateau.
For host particles ($\alpha={\rm H}$), $u_{\rm H}^2$ corresponds directly to the plateau height of $\gamma_{L,{\rm H}}(t)$.
For carrier particles ($\alpha={\rm C}$), the plateau is less pronounced due to stronger anharmonicity; accordingly, $u_{\rm C}^2$ is evaluated at the same $t_{\rm p}$ to represent their effective vibrational amplitude.
This definition enables a consistent comparison of Debye--Waller factors between the two species and provides a microscopic proxy for the local stiffness of their respective environments.
If $u_\alpha^2 \propto T$ holds, the dynamics in that regime can be regarded as effectively harmonic.


\subsection{Three-dimensional $\alpha$-AgI (FUH) model: interactions and simulation protocol}
To demonstrate the generality of our coarse-grained Coulombic framework, we perform molecular dynamics simulations of a 3D ionic conductor inspired by $\alpha$-AgI, following the Fukumoto--Ueda--Hiwatari (FUH) model~\cite{FukumotoJPSJ1982}. 

The interaction between ions $i$ and $j$ separated by a distance $r_{ij}$ is given by
\begin{equation}
\phi_{ij}(r_{ij}) = \epsilon \left( \frac{\sigma_i + \sigma_j}{r_{ij}} \right)^7 
+ \frac{C\, q_i q_j}{r_{ij}},
\end{equation}
which combines short-range repulsion with long-range Coulomb interactions. 
Here $q_i = z_i f$ denotes the dimensionless effective charge of ion $i$, where $z_i$ is the valency 
(e.g., $z_{\mathrm{Ag}} = +1$, $z_{\mathrm{I}} = -1$) and $f$ is the ionicity parameter. 
The Coulomb interaction is characterized by the prefactor 
$C = e^2/(4\pi \varepsilon_0) = 14.3996~\mathrm{eV}\,\text{\AA}$, 
which sets the electrostatic energy scale. In the simulations, this prefactor is absorbed into the chosen unit system 
(eV for energy and \text{\AA} for length).
This formulation is equivalent to the conventional expression $z_i z_j (f e)^2 / r_{ij}$ used in the original $\alpha$-AgI (FUH) model~\cite{FukumotoJPSJ1982}.
The long-range interactions are evaluated using Ewald summation under periodic boundary conditions.

The parameters used in the simulations are as follows. The energy scale is $\epsilon = 0.177~\mathrm{eV}$ with ionicity $f = 1.0$. 
The ionic radii are $\sigma_{\mathrm{I}} = 2.2~\text{\AA}$ for iodide and $\sigma_{\mathrm{Ag}} = 0.63~\text{\AA}$ for silver, 
giving effective pair diameters $\sigma_{\mathrm{I-I}} = 4.4~\text{\AA}$, $\sigma_{\mathrm{Ag-Ag}} = 1.26~\text{\AA}$, 
and $\sigma_{\mathrm{I-Ag}} = 2.83~\text{\AA}$.
A cutoff radius of $r_c = 12.0~\text{\AA}$ is used. The lattice constant is $a = 5.08~\text{\AA}$, and the system size corresponds to $N = 256$ particles.
The simulations are performed with a time step $\Delta t = 0.0093$ ps and a thermostat damping parameter $\tau = 0.1$ ps, over a total of $10^6$ steps. The system consists of Ag$^+$ and I$^-$ ions arranged in a body-centered cubic lattice with mobile silver ions occupying interstitial sites. 
Simulations are performed in the $NVT$ ensemble using a Nos\'e--Hoover thermostat and integrated via the velocity-Verlet algorithm. 

This model retains the essential ingredients of our two-dimensional system-long-range interactions, size asymmetry, and mobile charge carriers-demonstrating that the observed dynamical behavior is robust across dimensions.

\section{Acknowledgments}
The authors thank Prof.\ Junichi Kawamura for valuable discussions and Dr.\ Lukas Fischer for critical reading of the manuscript. The authors acknowledge support by the JST FOREST Program (grant no. JPMJFR212T, JPMJFR213H), AMED Moonshot Program (grant no. JP22zf0127009), JSPS KAKENHI (grant no. JP24H02203, JP24H02204), and Takeda Science Foundation.

\clearpage
\appendix
\setcounter{figure}{0}

\renewcommand{\thefigure}{S\arabic{figure}}
\section{SUPPORTING INFORMATION}

In the main study, we introduced a minimal two-dimensional binary model that isolates the microscopic origin of fast ion transport by disentangling sublattice melting, anharmonicity, and dynamical heterogeneity in a chemically agnostic framework. A rigid host lattice coexists with a soft carrier sublattice, producing a broad regime in which carriers delocalize while the host remains crystalline. Structural, dynamical, and time-correlation analyses demonstrate that fast transport in this regime arises from collective, anharmonic carrier motion rather than independent hopping.

While our study establishes the central phenomenology and its physical interpretation, several aspects merit further clarification and quantitative support. 
The Supporting Information addresses these points in detail and provides additional structural and dynamical diagnostics used to characterize sublattice order, collective carrier motion, and deviations from harmonic behavior across the density-controlled crossover discussed in the main text.
We first discuss issues related to the quantitative correspondence between simulation observables and experimental measures, providing additional context for interpreting diffusivity and structural metrics. We then examine the dimensional robustness of the sublattice melting and high-temperature transport behavior by extending the analysis to three-dimensional (3D) systems. The dependence of sublattice melting on carrier size in three dimensions is analyzed next, highlighting how steric confinement and migration barriers reshape the transport landscape. We further present a detailed analysis of the temperature evolution of the radial distribution function for the two-dimensional system at the packing fraction $\phi = 0.85$, clarifying the structural signatures associated with carrier delocalization. Finally, we investigate hyperuniformity and static density fluctuations through the structure factor, providing additional insight into long-wavelength correlations in the selectively melted state. In addition, we provide further analysis of lattice anharmonicity through the Lindemann index and Debye-Waller factor, together with a systematic examination of the density dependence of carrier dynamics and host-lattice fluctuations in the two-dimensional (2D) non-additive potential (NAP) model. We also clarify the geometric origin of the honeycomb-like carrier trajectories observed in the main text and discuss the connection between collective hopping, migration entropy, and effective activation barriers within a transition-state framework. The effects of partial carrier occupancy and the temperature dependence of spatial dynamical heterogeneity are further examined to assess the robustness of the collective transport mechanism identified in this work.

Together, these supporting analyses reinforce the conclusions of the main results, clarify the limits of the minimal model we proposed, and outline directions for connecting sublattice-melting physics to broader classes of superionic and glassy materials.

\subsection{Quantitative mapping between simulation and experiment}
\begin{figure}[b!]
    \centering
    \includegraphics[width=\linewidth]{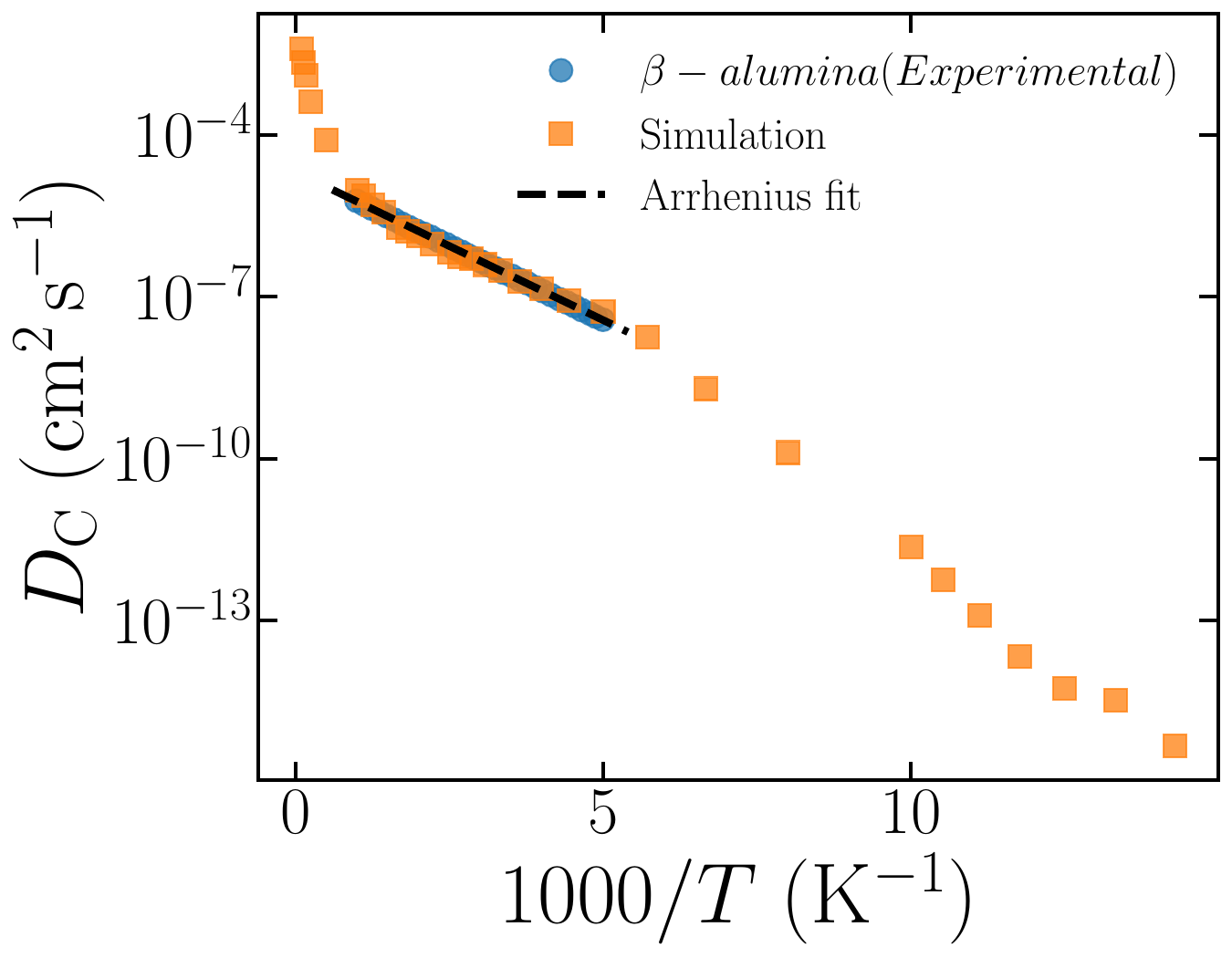}

    \caption{Collapse of simulated carrier diffusivity (orange squares) onto experimental $\beta$-alumina data (blue circles) in the Arrhenius regime. Simulation data are rescaled to experimental units. The black dotted line shows the common Arrhenius fit, demonstrating quantitative agreement between simulation and experiment.}
    \label{fig:fig1-supply_experimentalcomparison}
\end{figure}
To assess the quantitative validity of our minimal numerical model, we directly compared the carrier diffusivity obtained from two-dimensional simulations at packing fraction $\phi = 0.70$ with experimental diffusion data for $\beta$-alumina \cite{KamishimaJPSJ2010}, as shown in Fig.~\ref{fig:fig1-supply_experimentalcomparison}. The comparison was performed in the Arrhenius regime associated with sublattice melting, where both simulation and experiment exhibit approximately activated transport.

Our choice of $\beta$-alumina as a reference system is motivated by clear structural and dynamical parallels with our minimal NAP model. In both cases, mobile ions are constrained to a partially occupied sublattice, resulting in collective transport on a reduced-dimensional network \cite{KamishimaJPSJ2010}. In particular, the effective dimensionality and site occupancy in our two-dimensional simulations closely resemble those of Na-$\beta$-alumina conduction planes, where ionic motion is largely confined within well-defined layers.

At the microscopic level, Na-$\beta$-alumina exhibits multiple site types (e.g., BR and a-BR) with distinct local environments, which can introduce energetic heterogeneity and domain-like structures \cite{KamishimaJPSJ2010}. In contrast, our model assumes energetically equivalent sites, thereby isolating the essential transport physics from material-specific complexity. From this perspective, $\beta$-alumina serves as a representative sublattice-mediated ionic conductor, and the comparison is intended not for material-specific reproduction but to demonstrate that the collective, anharmonic transport mechanism identified in the simulations is consistent with experimental behavior.

Now, the experimental diffusivity follows an Arrhenius form,
\begin{equation}
D_{\mathrm{exp}}(T) = \frac{a^2}{\tau_0}\,\exp\!\left(-\frac{\epsilon}{k_\mathrm{B} T}\right),
\end{equation}
where $\epsilon$ is an effective activation energy and $a^2/\tau_0$ is a transport prefactor with dimensions of diffusivity. In the simulations, temperature and diffusivity are expressed in reduced units. To relate the two, we performed a direct collapse of the Arrhenius portions of the inverse-temperature plots by matching both the slope and intercept over the interval corresponding to the sublattice-melting regime.

This procedure yields an effective activation scale $\frac{\epsilon}{k_\mathrm{B}} = 1264.3~\mathrm{K}$, in quantitative agreement with reported activation energies for fast-ion conduction in $\beta$-alumina \cite{KamishimaJPSJ2010}. The vertical alignment of the Arrhenius plots further determines the transport prefactor $\frac{a^2}{\tau_0} = 2.05 \times 10^{-5}~\mathrm{cm}^2\,\mathrm{s}^{-1}$.

Importantly, only the ratio \ensuremath{a^2/\tau_0} is fixed by the collapse; separating the characteristic length scale \ensuremath{a} and the microscopic time scale \ensuremath{\tau_0} would require additional assumptions about single-particle hopping dynamics that lie outside the scope of the present minimal, chemically agnostic model. We therefore interpret \ensuremath{a^2/\tau_0} as an effective transport scale encoding collective, anharmonic carrier motion rather than independent activated hops.

This quantitative agreement demonstrates that the emergence of fast ion transport across sublattice melting in the simulations reproduces not only the qualitative phenomenology, but also the correct experimental activation scale and transport prefactor, supporting the relevance of collective and anharmonic mechanisms as the microscopic origin of superionic transport.

\subsection{Dimensional robustness and high-temperature transport behavior}
\noindent \textbf{Three-dimensional NAP model: simulation protocol.} To test the robustness of the observed behavior against dimensionality, we performed complementary 3D molecular dynamics simulations based on the same physical design principles as the 2D model. Initially the interaction hierarchy, size asymmetry, and mass ratios between host and carrier species were kept identical as the 2D NAP model (see ``Materials and Methods'' in main text), isolating dimensionality as the only difference.

The initial configuration was constructed on an FCC-like lattice with alternating host and carrier particles in all three directions. The simulation box comprised a periodic array of $4\times4\times4$ unit cells ($N\simeq512$ particles with equal species populations). Host particles had diameter $\sigma_{\mathrm{H}}=1.0$, carriers $\sigma_{\mathrm{C}}=0.154$, and both species had equal masses. The lattice spacing $a_{\mathrm{L\text{-}space}}=1.5$ corresponds to a volume packing fraction $\phi_\mathrm{volume}=0.6228$, yielding a cubic box of size $L=6.0$ with periodic boundary conditions.

Interactions followed the same scheme as in two dimensions: stiff, purely repulsive host-host and host-carrier interactions stabilized the host crystal, while softened carrier-carrier interactions promoted correlated carrier motion. After equilibration, microcanonical production runs were performed.\\ 
\begin{figure}[htbp!]
    \centering
    \includegraphics[width=\linewidth]{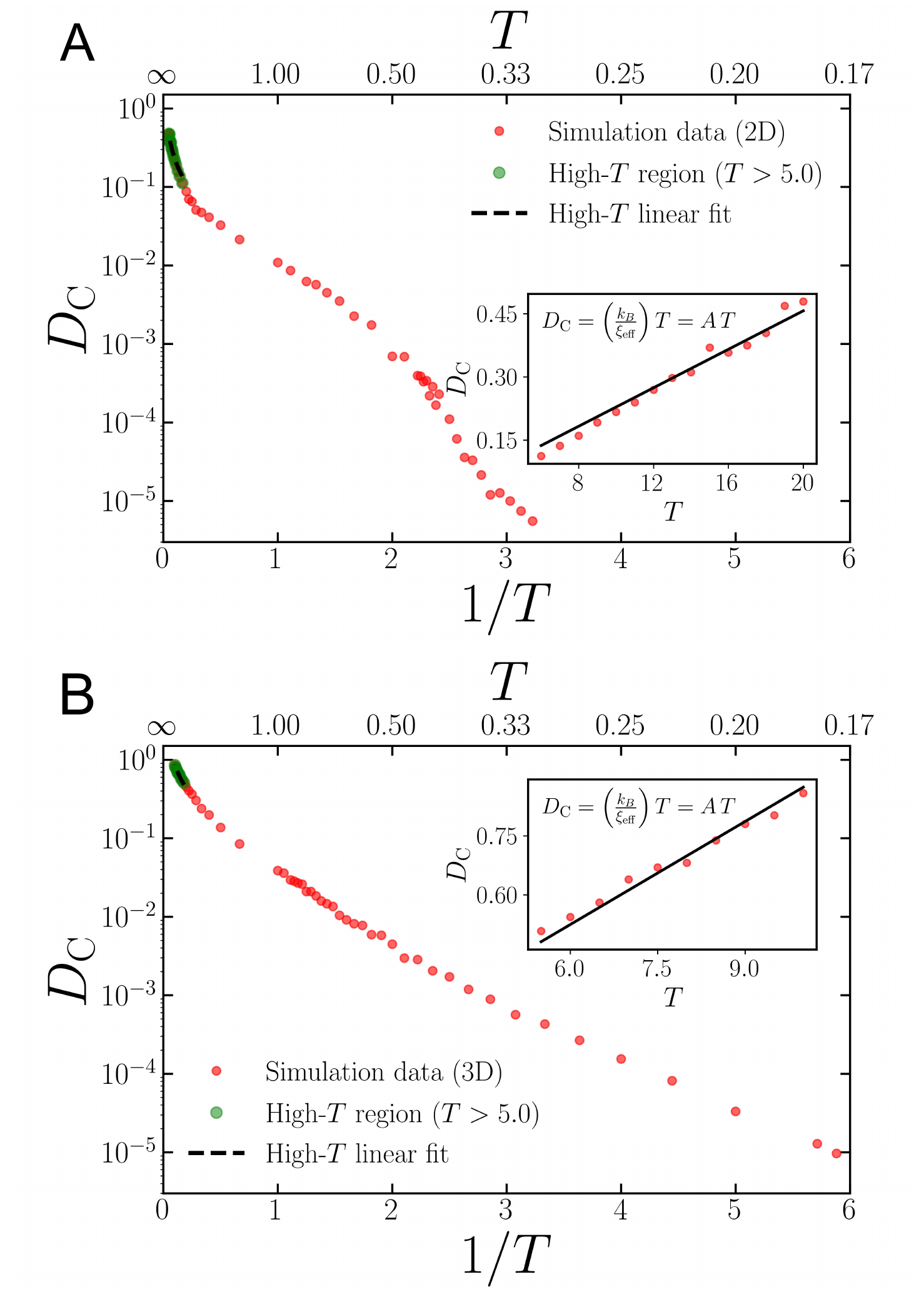}

    \caption{High-temperature transport of small carriers in two and three dimensions. (\textit{A}) Two-dimensional system at packing fraction $\phi = 0.85$ (similar to Fig.~2\textit{A} of the main text). (\textit{B}) Three-dimensional system at volume fraction $\phi_\mathrm{volume} = 0.62$. In both panels, the main plots show the carrier diffusivity $D_\mathrm{C}$ as a function of inverse temperature $1/T$. A distinct high-temperature regime emerges in which the diffusivity exhibits weak curvature in the Arrhenius representation. The shaded (green) symbols highlight the high-temperature region used for further analysis. Insets show $D$ plotted directly as a function of $T$ for this high-temperature regime, together with linear fits of the form $D_\mathrm{C} = \left(k_\mathrm{B} / {\xi_{\mathrm{eff}}} \right) T = A\,T$. The excellent linearity observed in both two and three dimensions confirms the validity of the Einstein relation in this regime, indicating a crossover to a kinetic, weakly constrained transport mechanism at elevated temperatures.}
    \label{fig:fig2-supply_2d-3D}
\end{figure}

\noindent \textbf{High-temperature transport of carriers in 2D and 3D systems.} We first characterize the high-temperature transport regime and its dependence on dimensionality for the small-carrier system (carrier size $\sigma_\mathrm{C} = 0.154$), where steric constraints are minimal. Figure~\ref{fig:fig2-supply_2d-3D} compares the carrier diffusivity $D$ as a function of inverse temperature $1/T$ in two dimensions at packing fraction $\phi = 0.85$ (Fig.~\ref{fig:fig2-supply_2d-3D}\textit{A}) and in three dimensions at volume fraction $\phi_\mathrm{volume} = 0.62$ (Fig.~\ref{fig:fig2-supply_2d-3D}\textit{B}), using identical interaction parameters and carrier size (see ``Materials and Methods'' in main text).

In both cases, the diffusivity initially follows an activated, Arrhenius-like behavior at low temperatures, reflecting carrier motion constrained by the crystalline host lattice. Upon increasing temperature, the system enters an intermediate sublattice-melting regime in which carrier mobility increases rapidly while the host lattice remains largely intact. In three dimensions, the additional configurational freedom smooths this low-temperature crossover, such that the deeply activated regime evident in two dimensions is compressed to lower temperatures and not fully resolved within the accessible simulation window. Following full melting of the host lattice, the diffusivity does not saturate, but instead crosses over into a distinct high-temperature transport regime characterized by a nearly linear increase of diffusivity $D$ with temperature $T$.

This crossover is made explicit in the insets of Fig.~\ref{fig:fig2-supply_2d-3D}, where $D$ is plotted directly as a function of $T$. In this high temperature regime, the diffusivity is well described by $D = \left(k_\mathrm{B} / {\xi_{\mathrm{eff}}} \right) T = A\,T$, consistent with the Einstein relation for kinetic, weakly constrained transport. Linear fits yield $A = k_\mathrm{B} / {\xi_{\mathrm{eff}}} = 2.29 \times 10^{-2}$ for the two-dimensional system and $A = k_\mathrm{B} / {\xi_{\mathrm{eff}}} = 8.74 \times 10^{-2}$ for the three-dimensional system. The robustness of this linear scaling across dimensionality demonstrates that, once structural constraints are sufficiently relaxed, carrier motion is governed primarily by thermal energy rather than by activation over persistent barriers. Notably, the same sequence of transport regimes-activated at low T, an intermediate sublattice-melting crossover, and the high-temperature linear $D \propto T$ behavior-is observed in three dimensions, confirming that the phenomenology is not an artifact of two-dimensionality.

This behavior stands in clear contrast to that observed for large carriers (discussed in the next section), where the diffusivity develops a weakly temperature-dependent, saturation-like regime at high temperatures. In those systems, transport remains dominated by residual activation barriers even after host melting, leading to an approximately temperature-independent diffusivity. The absence of such saturation in the small-carrier system highlights a qualitative change in the dominant transport mechanism, underscoring the role of carrier size in determining whether the high-temperature dynamics are barrier-limited or kinetic in nature.

\subsection{Carrier-size dependence of sublattice melting in three dimensions}
\begin{figure}[t!]
    \centering
    \includegraphics[width=\linewidth]{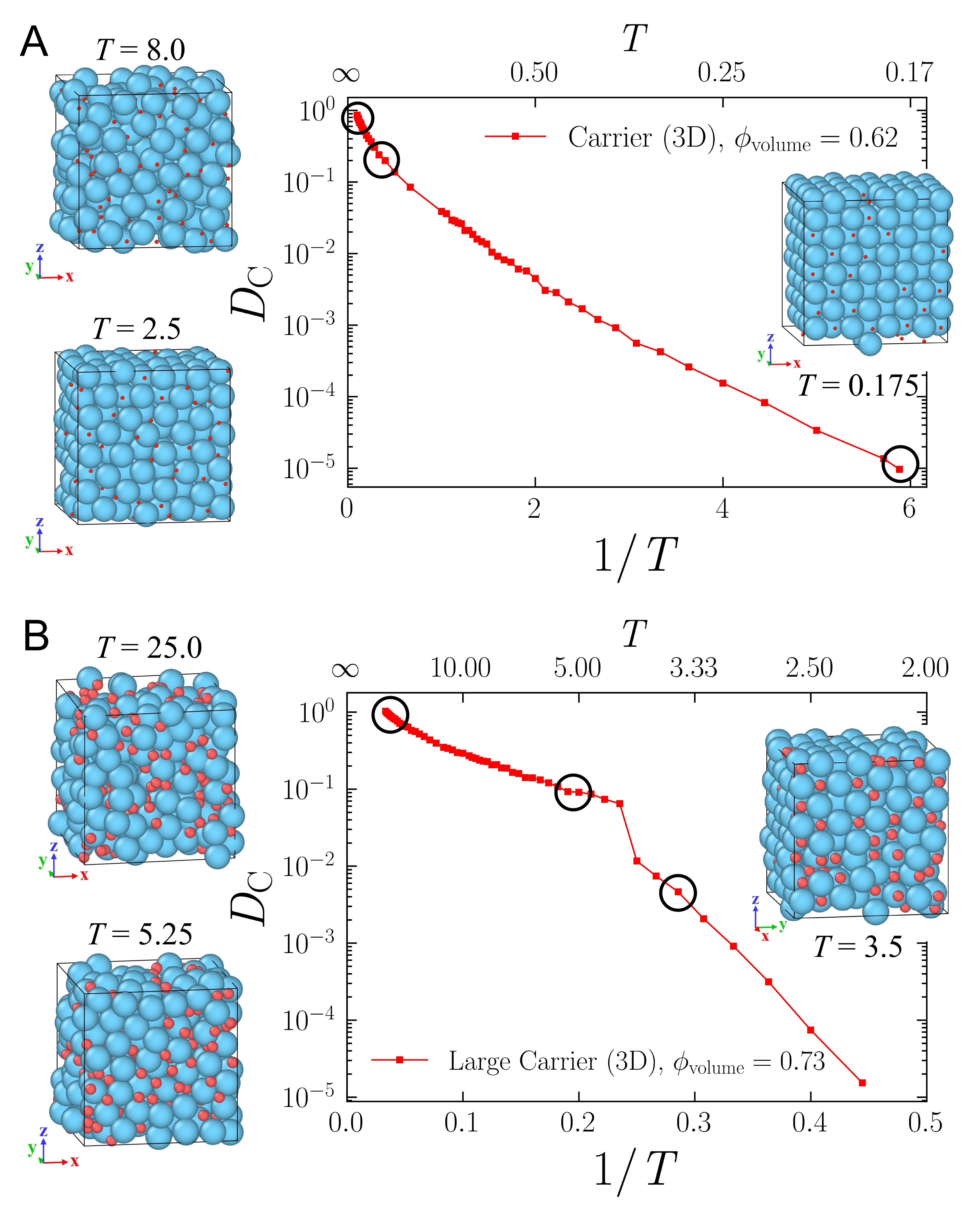}

    \caption{Carrier diffusivity $D_\mathrm{C}$ as a function of inverse temperature $1/T$ in the three-dimensional system for two carrier sizes. (\textit{A}) Small carriers show high mobility and a broad sublattice-melting regime while the host lattice remains crystalline (same as Fig.~\ref{fig:fig2-supply_2d-3D}\textit{B}). Snapshots at $T=0.175$ reveal liquid-like carrier motion within an ordered host, at $T=2.5$ a dynamically disordered, percolating carrier network with intact host order, and at $T=8.0$ complete melting of both sublattices. (\textit{B}) Larger carriers exhibit a sharp onset of diffusion near host-lattice melting. Configurations at $T=3.5$ indicate carrier sublattice melting, at $T=5.25$ signatures of partial host-lattice destabilization, and at $T=25.0$ full melting of the crystal. Black circles mark the corresponding temperatures.}
    \label{fig:fig3-supply_3D}
\end{figure}
\begin{figure*}[t!]
    \centering
    \includegraphics[width=\linewidth]{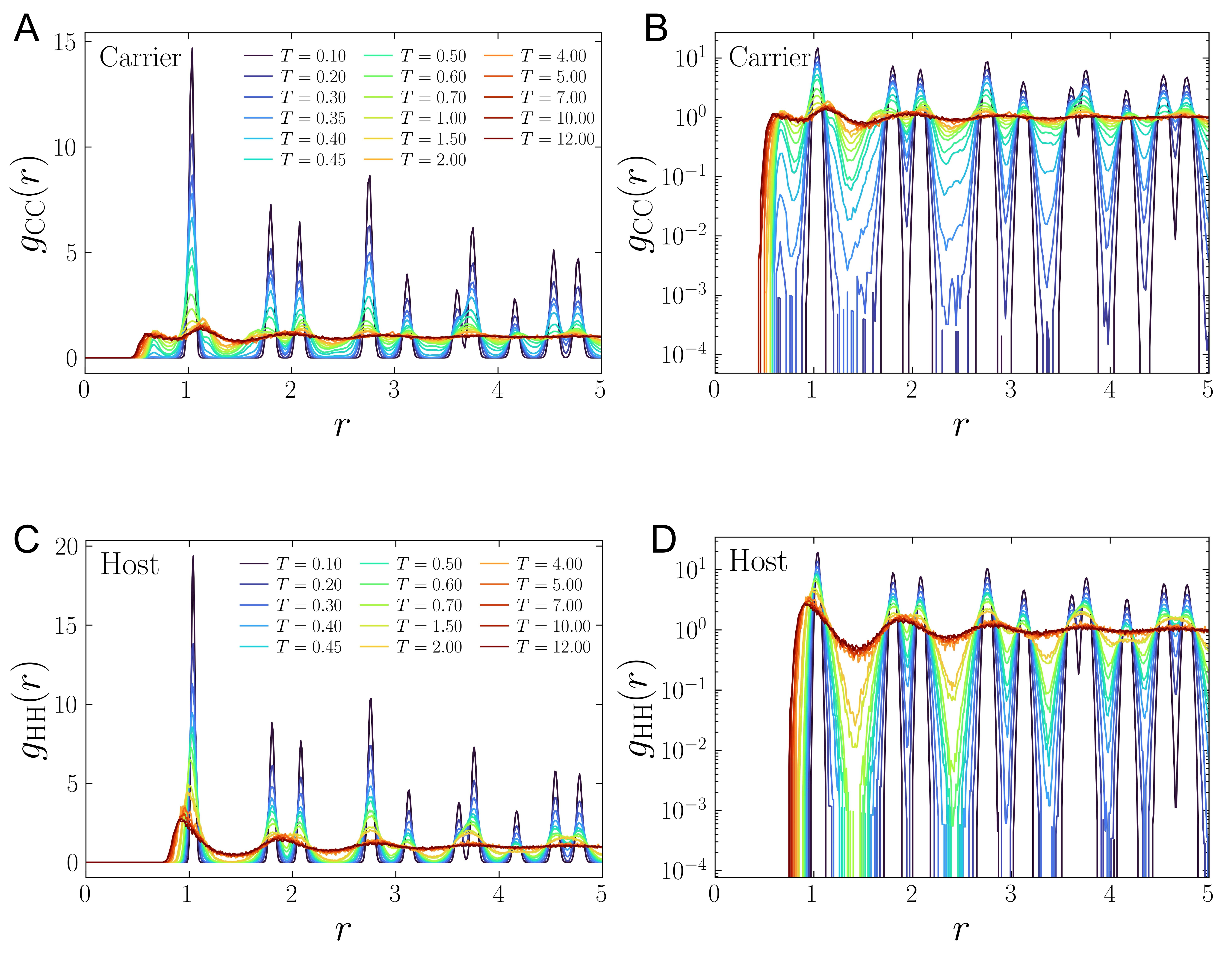}

    \caption{Temperature dependence of the radial distribution function $g_{\alpha\alpha}(r)$ for the two-dimensional system at packing fraction $\phi = 0.85$. (\textit{A--B}) Carrier ($\alpha$: C) distribution $g_{\mathrm{CC}}(r)$ at low and elevated temperatures, showing sharp crystalline peaks at low $T$ and the emergence of an excess shoulder at higher $T$. (\textit{C--D}) Host ($\alpha$: H) distribution $g_{\mathrm{HH}}(r)$, which remains largely ordered until complete melting, with negligible excess peaks or shifts. The logarithmic representation (\textit{B, D}) highlights subtle features of the distribution.}
    \label{fig:fig4_supply_RDF}
\end{figure*}
We examine how carrier size controls the emergence and character of sublattice melting in three dimensions by comparing transport and structural signatures for different carrier diameters at fixed lattice spacing $a_\mathrm{L-space} = 1.5$.

Figure~\ref{fig:fig3-supply_3D}\textit{A} shows the carrier diffusivity $D$ as a function of inverse temperature $1/T$ for small carriers with $\sigma_{\mathrm{C}} = 0.154$. The diffusivity exhibits a broad intermediate temperature regime in which $D$ remains finite and only weakly temperature dependent over an extended interval. This behavior signals a wide sublattice-melting window, during which carriers progressively delocalize and form a dynamically connected network while the host lattice retains long-range crystalline order. The persistence of finite diffusivity throughout this regime indicates that carrier motion is not governed by a single dominant activation barrier, but instead reflects a distribution of low-energy migration pathways enabled by the size mismatch between carriers and the host lattice. Notably, this phenomenology closely mirrors that observed in two dimensions (see previous section), demonstrating that the emergence of an extended sublattice-melting regime for small carriers is robust to dimensionality.

Representative configurations at selected temperatures within this regime, marked by black circles in Fig.~\ref{fig:fig3-supply_3D}\textit{A}, provide direct structural support for this interpretation. At $T = 0.175$, carriers already display liquid-like mobility despite a fully ordered host lattice, indicating that sublattice melting sets in deep in the low-temperature regime. At $T = 2.5$, the host lattice remains intact while the carrier subsystem forms a dynamically disordered, percolating structure. Only at $T = 8.0$ do both sublattices lose structural order, marking complete melting of the crystal.

In contrast, for larger carriers with $\sigma_{\mathrm{C}} = 0.42$ (Fig.~\ref{fig:fig3-supply_3D}\textit{B}), strong steric confinement within the host lattice leads to substantially enhanced effective migration barriers. Carrier mobility remains strongly suppressed until the host lattice begins to lose structural integrity, producing a sharp onset of diffusion near the host melting temperature, followed by a weakly temperature-dependent, saturation-like diffusivity at higher temperatures. Notably, this combination of an abrupt activation of mobility and subsequent diffusivity saturation closely resembles the transport signatures reported in several superionic conductors \cite{TakeiriNatMater2022}. 
Structural analysis confirms that this apparent saturation-like behavior corresponds to a regime of \emph{partial host melting}, in which ordered and disordered host regions coexist and continue to impose geometric constraints on carrier motion (see the snapshot at $T = 5.25$). This behavior can be rationalized by an activated form, $D \sim \exp\!\left(-\Delta E / k_{\mathrm{B}}T\right)$, which at sufficiently high temperatures may be expanded as $\exp\!\left(-\Delta E / k_{\mathrm{B}}T\right) \simeq 1 - \Delta E / k_{\mathrm{B}}T$, yielding only a weak residual temperature dependence. Frequent collisions and persistent geometric constraints within the partially disordered matrix further suppress the growth of diffusivity, leading to the observed saturation-like behavior. 

Upon further heating, the diffusivity exhibits an additional sharp increase, signaling complete melting of the host lattice (see the snapshots at $T = 25.0$ showing the complete melting scenario) and the disappearance of persistent steric constraints. In this high-temperature limit, the system crosses over to a liquid-like transport regime in which carrier motion becomes weakly correlated and predominantly governed by thermal energy. Consistent with previous observations, the diffusivity in this regime follows an Einstein-like scaling $D \propto k_{\mathrm{B}}T$, indicating that transport is no longer controlled by activation over geometric barriers but by kinetic motion in a fully disordered environment. We note that this ultimate high-temperature crossover, while clearly resolved in simulations, is expected to be difficult to access experimentally. The temperatures required to fully eliminate residual steric constraints typically exceed those attainable in stable solid-state or superionic materials, where thermal decomposition or chemical degradation intervene. As a result, experimental measurements often probe only the intermediate, weakly temperature-dependent regime, which may appear as an apparent saturation of diffusivity.

Together, these results demonstrate that the carrier size acts as a key control parameter governing both the width of the sublattice-melting regime and the dominant transport mechanism at high temperatures. Small carriers transition to a kinetic, Einstein-like transport regime once structural constraints are lifted, whereas large carriers remain effectively barrier-limited even in the disordered state. Similar qualitative behavior can also be induced in two dimensions by tuning model parameters.

\subsection{Static property: Radial distribution function (RDF), $g_{\alpha\alpha}(r)$ of the system}
\begin{figure}[b!]
    \centering
    \includegraphics[width=\linewidth]{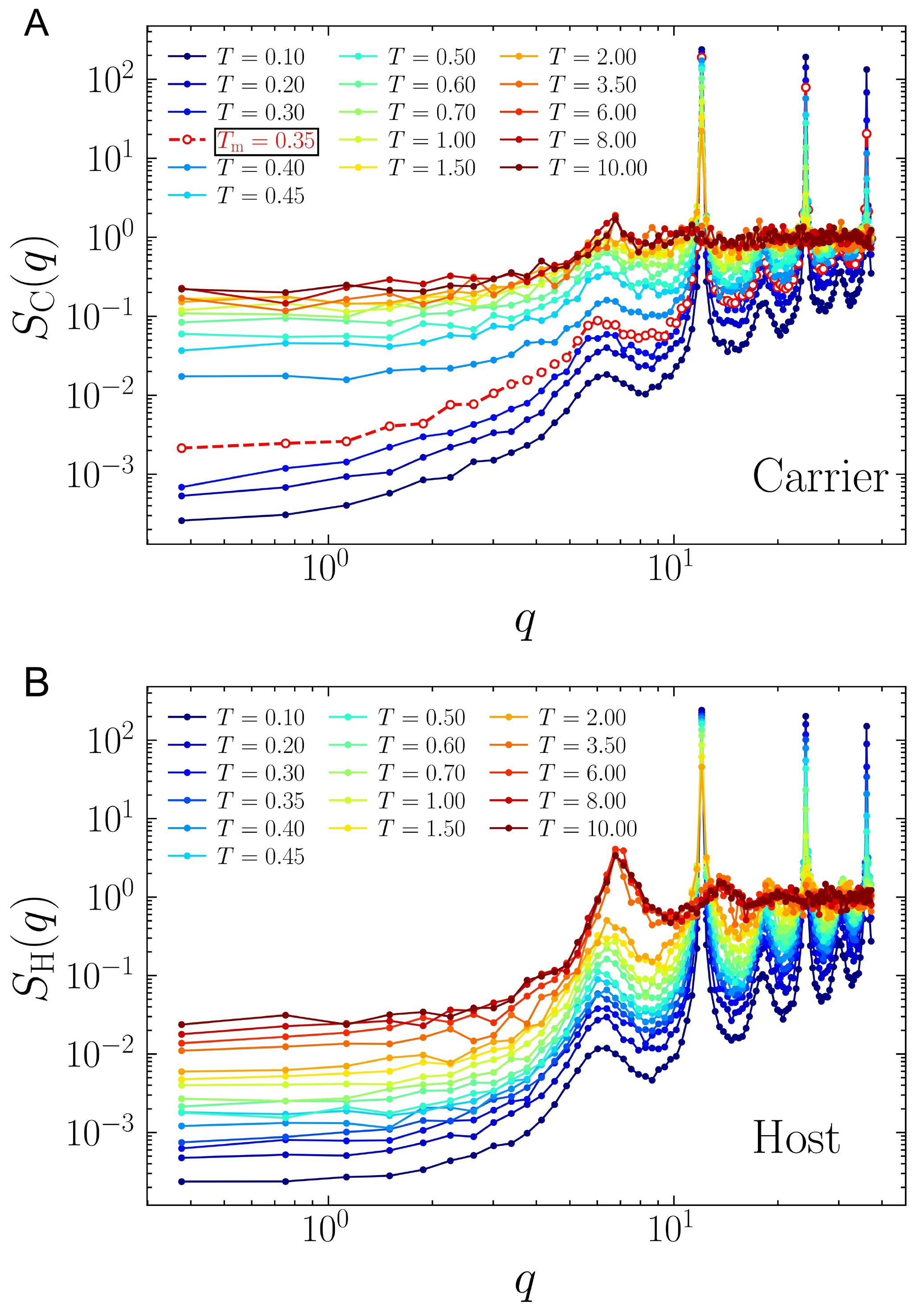}

    \caption{Static structure factor $S_\alpha(q)$ for the two-dimensional system at packing fraction $\phi = 0.85$. (\textit{A}) Carrier particles and (\textit{B}) host particles at representative temperatures across the sublattice-melting regime. Carriers exhibit strong small-$q$ suppression at low temperatures, which disappears upon melting ($T > T_\mathrm{m}$), whereas the host lattice retains pronounced Bragg-like order over a broader temperature range.}
    \label{fig:fig5-supply_sq}
\end{figure}
\begin{figure*}[t!]
    \centering
    \includegraphics[width=\linewidth]{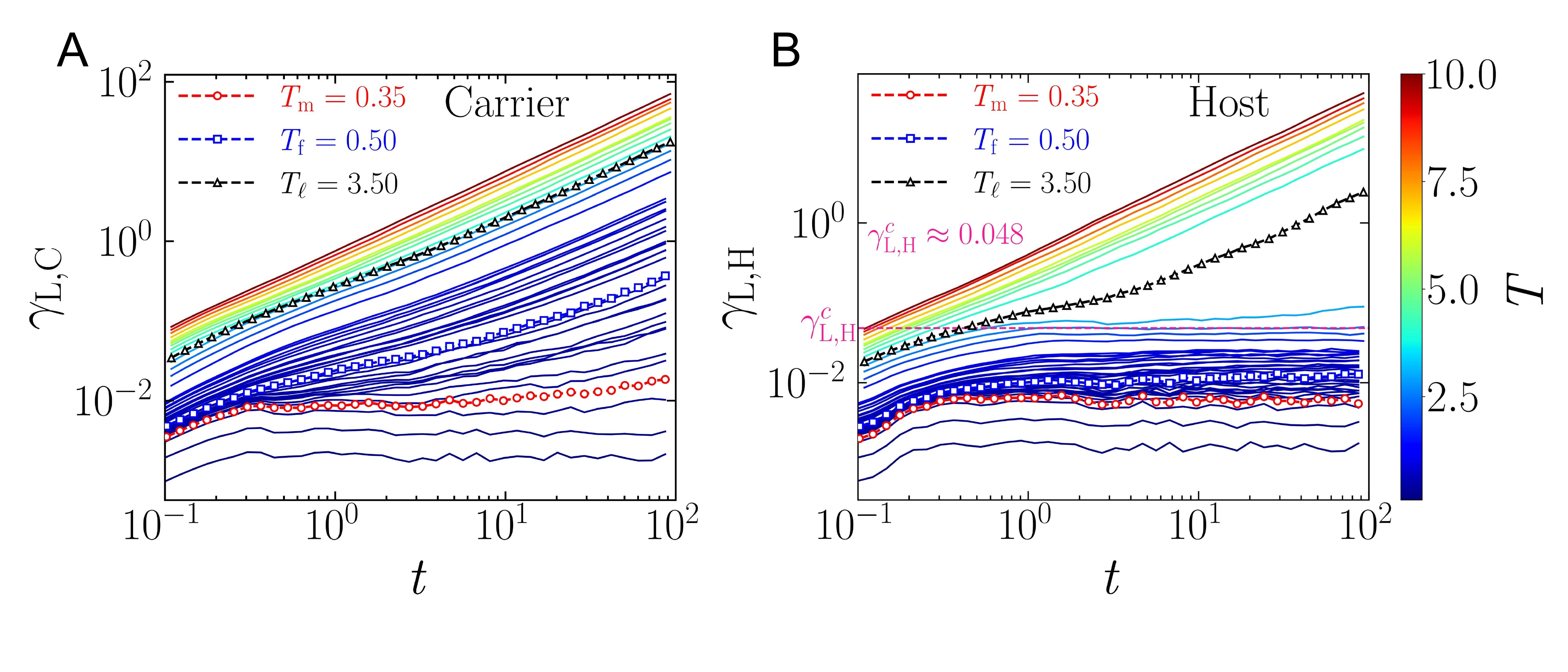}
    \caption{%
        Lindemann index $\gamma_{L,\alpha}$ and effect of anharmonicity: (\textit{A--B}) Time evolution of the Lindemann index for carriers and hosts at packing fraction $\phi = 0.85$. For the host lattice (\textit{A}), the Lindemann index shows a clear signature of lattice melting, while for the carriers (\textit{B}) the growth is purely driven by anharmonic fluctuations rather than a melting transition. 
    }
    \label{fig:fig6_supply_LindemannIndex}
\end{figure*}
In this section, we next analyse the temperature evolution of the radial distribution function (see ``Materials and Methods'' in the main text) for carriers $g_{\mathrm{CC}}(r)$ and hosts $g_{\mathrm{HH}}(r)$ separately. As shown in Fig.~\ref{fig:fig4_supply_RDF}\textit{A}, the carrier distribution at low temperatures exhibits sharp, periodic peaks, characteristic of crystalline order. Upon heating, a distinct shoulder emerges on the left of the first peak and gradually intensifies, accompanied by peak broadening and shifts near the sublattice-melting regime. The appearance of this excess feature reflects activated single-carrier hopping events associated with Frenkel-type disorder \cite{FunkeSTAM2013} (see Fig.~\ref{fig:fig4_supply_RDF}\textit{B} for its enhanced visibility on a logarithmic scale). With increasing temperature, these structural signatures indicate that the carrier sublattice progressively loses positional order and acquires liquid-like characteristics, while the host lattice remains largely intact. In contrast, the host distribution (Fig.~\ref{fig:fig4_supply_RDF}\textit{C--D}) shows neither the excess peak nor noticeable shifts until its complete melting. This clear disparity highlights the heterogeneous nature of the system, where mobile carriers undergo an early transition to a fluid-like state within an otherwise ordered host framework.  

\subsection{Hyperuniformity: Structure factor}
We analyze the structure factor
\begin{equation}
  S_\alpha(q) = \frac{1}{N_\alpha}\,\big\langle \rho_\alpha(\mathbf{q})\,\rho_\alpha(-\mathbf{q}) \big\rangle
\end{equation}
for both the carrier and host lattices at different temperatures (see Fig.~\ref{fig:fig5-supply_sq}), where $N_\alpha$ is the number of particles of species $\alpha$, $\rho_\alpha(\mathbf{q})=\sum_{j\in\alpha} e^{i\mathbf{q}\cdot\mathbf{r}_j}$ is the Fourier component of the density, and $\langle\cdots\rangle$ denotes an ensemble average.
At low temperatures, the carrier particles exhibit strong hyperuniformity, reflected in a vanishing $S_\alpha(q \to 0)$, which indicates suppressed long-wavelength density fluctuations and persistent crystalline correlations \cite{TorquatoPhysRep2018}.
Such hyperuniformity is also characteristic of ordered systems such as Wigner crystals, consistent with the quasi-crystalline arrangement of the carrier sublattice observed here.
As the system approaches the sublattice melting regime, $S_\alpha(q)$ for carriers develops pronounced fluctuations and the small-$q$ suppression disappears, signaling the breakdown of hyperuniformity and the loss of long-range crystalline correlations.
Microscopically, the loss of hyperuniformity is localized near the temperature $T_\mathrm{m}$ (see the highlighted red circles in Fig.~\ref{fig:fig5-supply_sq}\textit{A}). For $T<T_\mathrm{m}$, the carrier sublattice remains hyperuniform, consistent with a Wigner-crystal-like state in which long-wavelength density fluctuations are suppressed. Near $T_\mathrm{m}$, carriers increasingly occupy heterogeneous, interstitial configurations, resulting in enhanced long-wavelength density fluctuations and the breakdown of hyperuniformity. This real-space heterogeneity is reflected in pronounced spatial variations of local carrier environments.
This transition reflects the fragility of the carrier sublattice: once thermal fluctuations overcome the ordering, the carriers rapidly become disordered.
Even after complete melting, the carrier $S_\mathrm{C}(q)$ exhibits only moderate peak heights, consistent with a fluid-like state lacking strong residual correlations.

In contrast, the host lattice maintains structural order over a broader temperature range. Although hyperuniformity is weaker for hosts at low temperatures, a sharp Bragg-like peak emerges and grows significantly after the host lattice completes melting, reflecting the development of long-range positional correlations as the host particles reorganize. These observations highlight the asymmetric melting dynamics between carriers and hosts, and the sensitivity of hyperuniformity to sublattice destabilization.
\begin{figure*}[htbp!]
    \centering
    \includegraphics[width=\linewidth]{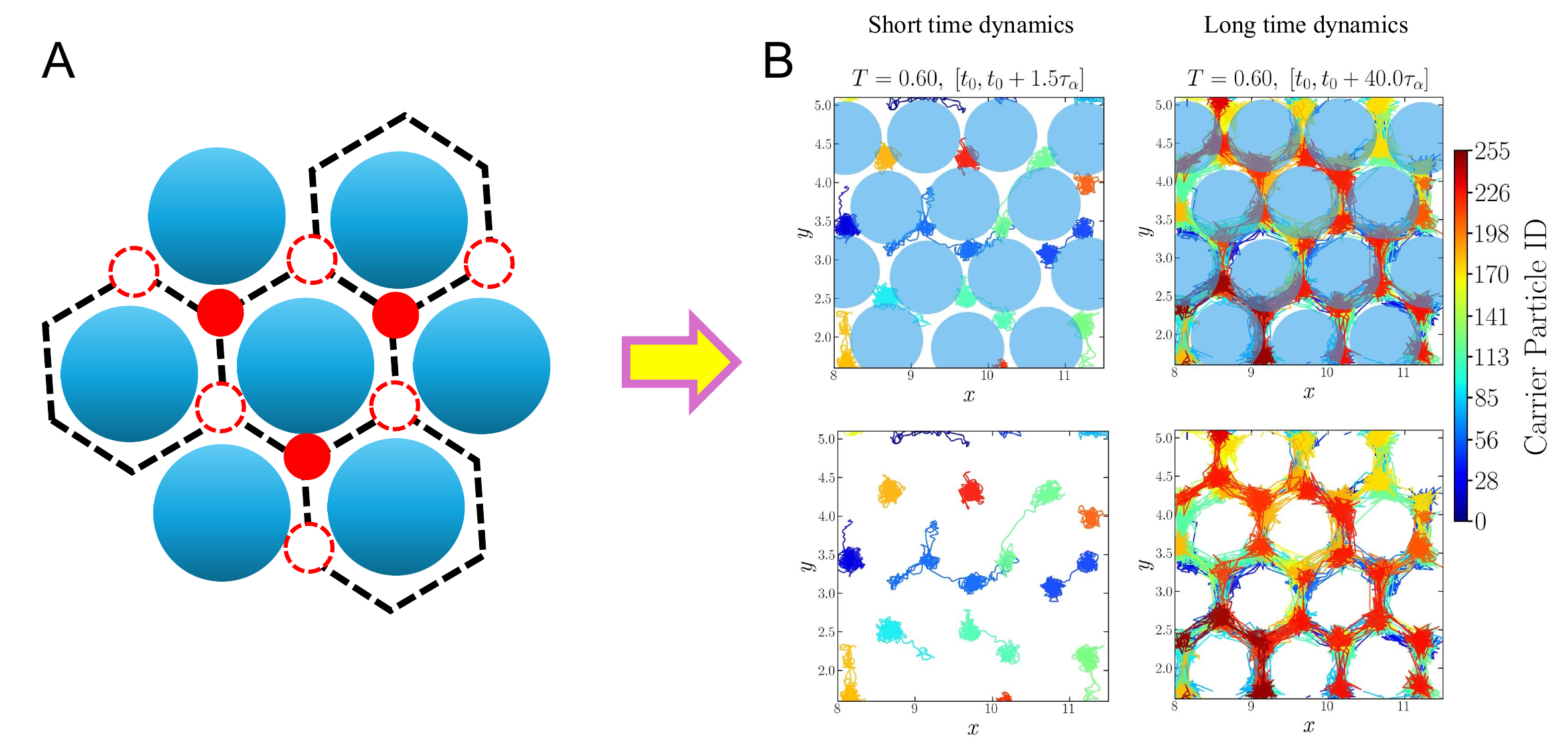}
    \caption{Carrier trajectories and interstitial network structure: Trajectory of carrier particles forming hexagonal-like structures. (\textit{A}) Schematic illustration: cyan particles represent host particles and red particles represent carriers. The dashed honeycomb network indicates the expected connectivity of interstitial pathways, and the dashed circular regions denote interstitial sites accessible to carriers. (\textit{B}) Actual carrier trajectories overlaid on host particle positions. Colors of the trajectories correspond to particle IDs, following the same convention as in Fig.~3 in the main text. Importantly, carrier motion is confined to the interstitial network and does not pass through host particle cores.}
    \label{fig:fig7_supply_trajectory}
\end{figure*}
\subsection{Lindemann Index Calculation}
The evolution of the Lindemann index (see ``Materials and Methods'' of the main text) reveals a pronounced enhancement of anharmonicity as the system approaches the sublattice melting point (Fig.~\ref{fig:fig6_supply_LindemannIndex}). Figs.~\ref{fig:fig6_supply_LindemannIndex}\textit{A,B} show the time evolution of the Lindemann index, $\gamma_{{\rm L},\alpha}(t)$, for the host and carrier particles at packing fraction $\phi = 0.85$. For the host lattice, $\gamma_{{\rm L},{\rm H}}(t)$ exhibits a well-defined plateau reflecting harmonic vibrations within a stable crystal, followed by a rapid increase once $\gamma_{{\rm L},{\rm H}}(t)$ exceeds the critical value $\gamma_{{\rm L},{\rm H}}^{\rm c} \simeq 0.048$, consistent with the conventional Lindemann criterion for lattice melting and close to previously reported critical values $\gamma_{\rm L}^{\rm c} \simeq 0.03$ for crystalline solids~\cite{ZahnPRL1999}.
In contrast, $\gamma_{{\rm L},{\rm C}}(t)$ for the carriers increases smoothly without a sharp transition, indicating that their fluctuations are governed by intrinsic anharmonic vibrations rather than by a structural instability.

\subsection{Origin of honeycomb-like carrier trajectories}
To clarify the origin of the honeycomb-like trajectories observed in Fig.~3 of the main text, we note that these patterns do not indicate hopping through host sites. Instead, they reflect the geometry of the interstitial network defined by the host lattice, together with the finite temporal resolution of trajectory sampling.

As shown in Fig.~\ref{fig:fig7_supply_trajectory}\textit{A}, the host particles form a hexagonal lattice. The interstitial voids define a connected network of accessible sites for the carriers, with pathways (indicated by black dotted lines) that are topologically equivalent to a honeycomb lattice. Carrier motion is therefore confined to these interstitial channels and naturally produces honeycomb-like trajectory patterns, while host cores remain excluded.

The apparent straight segments arise from the finite sampling interval. For example, in the case of $T = 0.60$, configurations are recorded every 500 steps (short-time) and 3900 steps (long-time), so intermediate positions are not resolved, leading to visual interpolation across excluded regions.

To clarify this point, Fig.~\ref{fig:fig7_supply_trajectory}\textit{B} shows a zoomed-in view with host particles overlaid (upper panel) and carrier trajectories alone (lower panel). Host particles are shown at their reference positions; their thermal displacements are small and do not alter the connectivity of the interstitial network. These plots confirm that carriers remain confined to interstitial regions and do not pass through host sites.
\begin{figure}[!htbp]
    \centering
    \includegraphics[width=\linewidth]{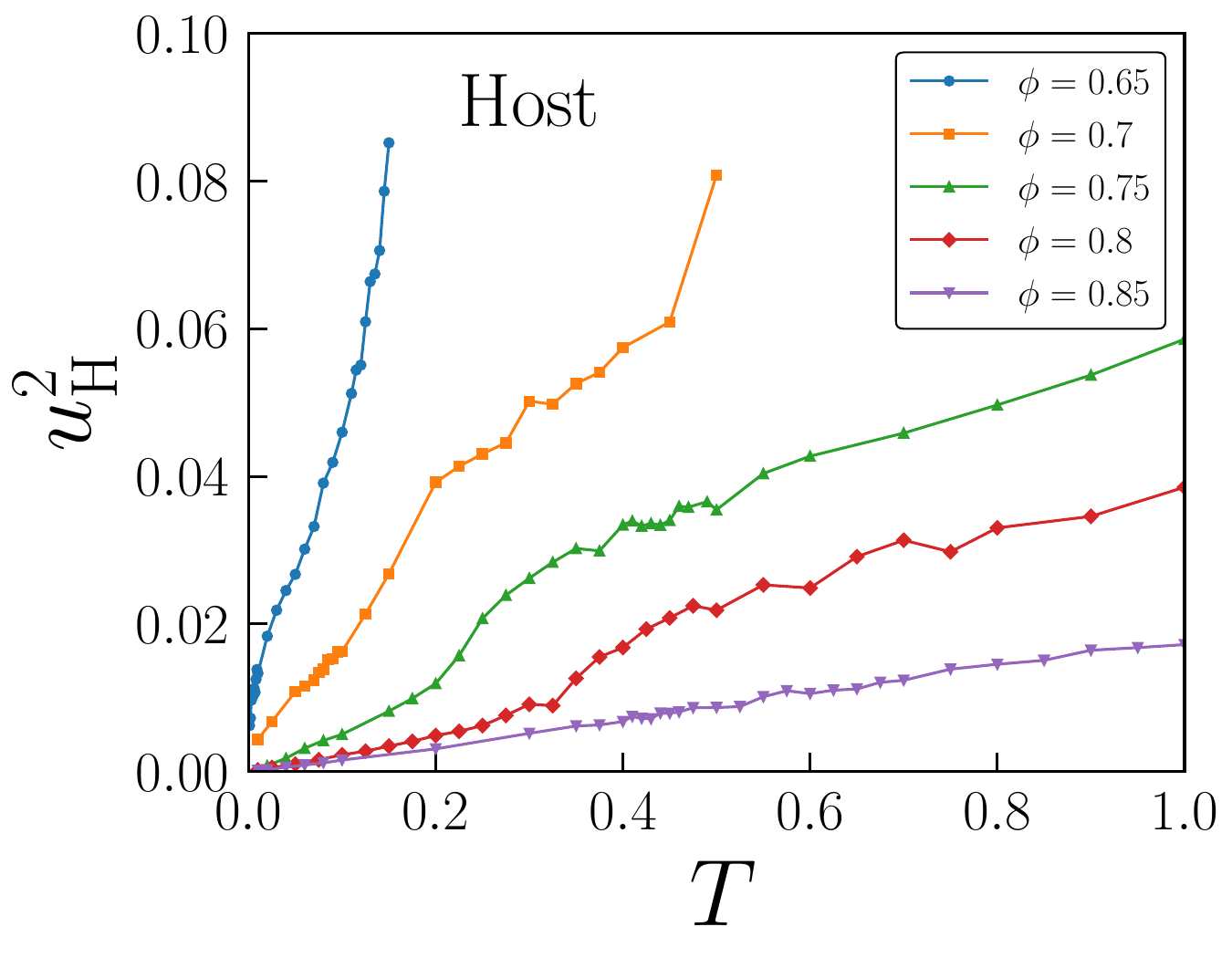}
   \caption{Density dependence of the host Debye--Waller factor (DWF) as a function of temperature $T$. While the high-density case ($\phi=0.85$) shows nearly linear (harmonic-like) behavior (same as Fig.~6 of the main text), lower-density systems display pronounced deviations from linearity, signaling the onset of anharmonic lattice dynamics associated with lattice softening.}
    \label{fig:fig8_supply_dwf_host}
\end{figure}
\begin{figure*}[!t]
    \centering
    \includegraphics[width=\linewidth]{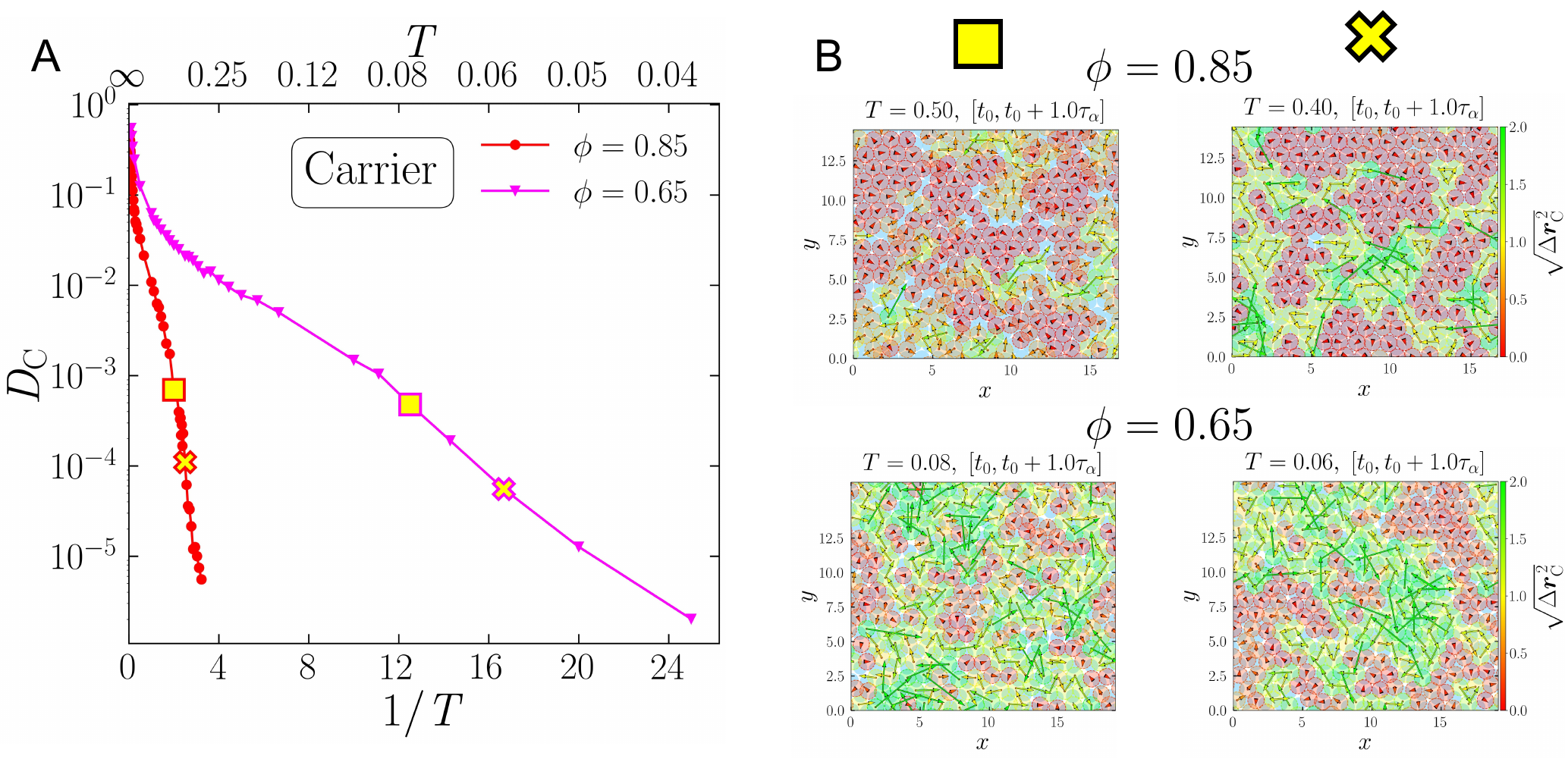}
    \caption{Density dependence of carrier dynamics. (\textit{A}) Carrier diffusivity ($D_\mathrm{C}$) as a function of inverse temperature ($1/T$) for $\phi = 0.65$ and $0.85$, highlighting the shift in the onset of sublattice melting. (\textit{B}) Spatial maps of carrier displacements $\sqrt{\Delta \boldsymbol{r}_\mathrm{C}^{\,2}}$ over $[t_0, t_0 + \tau_\alpha]$ near the onset: for $\phi = 0.85$ ($T = 0.50, 0.40$), dynamics are strongly heterogeneous with localized, string-like cooperative motion; for $\phi = 0.65$ ($T = 0.08, 0.06$), mobility is more spatially extended with weaker heterogeneity due to increased lattice softness.}
    \label{fig:fig9_supply_density_spatialMap}
\end{figure*}

\subsection{Density dependence of lattice anharmonicity and carrier dynamics in our 2D NAP model}
To elucidate the role of density in controlling lattice dynamics and carrier transport, we perform a systematic analysis across different packing fractions. We focus first on the Debye-Waller factor (DWF) of the host lattice as a measure of vibrational amplitude and lattice stiffness.

As shown in Fig.~\ref{fig:fig8_supply_dwf_host}, the density dependence of the DWF reveals a clear change in behavior. At lower density ($\phi = 0.65$), the host exhibits an early deviation from linear temperature dependence as the system approaches the melting transition, indicating enhanced lattice softness and increased anharmonicity. In contrast, at higher density ($\phi = 0.85$), the DWF remains approximately linear over a wider temperature range, consistent with a more rigid lattice.

To further elucidate how this difference in lattice response affects carrier dynamics, we analyze spatial maps of particle mobility near the onset of sublattice melting (Fig.~\ref{fig:fig9_supply_density_spatialMap}). At high density ($\phi = 0.85$), the dynamics are strongly heterogeneous and spatially localized, with pronounced string-like cooperative motion. These correlated displacements lead to the formation of extended immobile regions, reflecting constrained transport through a rigid environment. In contrast, at lower density ($\phi = 0.65$), the softer lattice supports more spatially extended mobility. Here, slow domains become broader while immobile regions shrink, resulting in more liquid-like and less spatially localized dynamics with weaker cooperative character.

\begin{figure*}[!t]
    \centering
    \includegraphics[width=\linewidth]{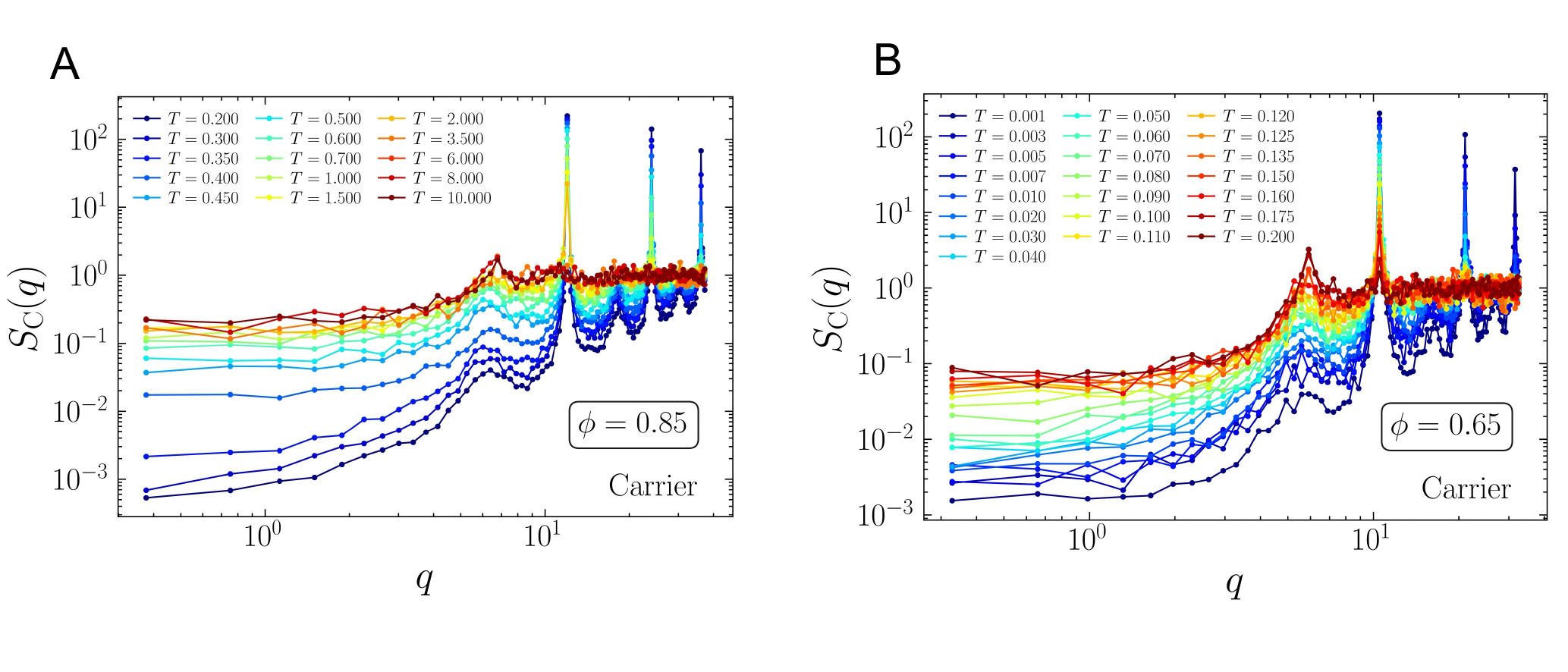}
    \caption{Density dependence of the static structure factor $S_{\mathrm{C}}(q)$ of the carrier particles at different temperatures $T$. (A) At higher density ($\phi = 0.85$), a pronounced suppression of $S_{\mathrm{C}}(q)$ in the low-$q$ regime at low $T$ indicates strong hyperuniform behavior. (B) At lower density ($\phi = 0.65$), $S_{\mathrm{C}}(q)$ remains finite at small $q$ even at very low $T$, indicating the absence of hyperuniformity.}
    \label{fig:fig10_supply_density_structureFactor}
\end{figure*}
In addition, we performed a structure factor analysis to quantify long-wavelength density fluctuations across densities (Fig.~\ref{fig:fig10_supply_density_structureFactor}). At high density ($\phi = 0.85$), the carrier structure factor exhibits $S_\mathrm{C}(q) \to 0$ as $q \to 0$ at low temperatures, indicating a strong tendency toward hyperuniformity and suppressed large-scale density fluctuations. In contrast, at lower density ($\phi = 0.65$), $S_\mathrm{C}(q)$ approaches a finite constant in the same limit, demonstrating the absence of hyperuniformity and significantly enhanced density fluctuations.

These reciprocal-space results are fully consistent with the real-space observations: higher density enforces stronger collective constraints, leading to localized, string-like motion and pronounced dynamical heterogeneity, whereas lower density promotes broader, less correlated mobility and a more gradual crossover in dynamical behavior.

Importantly, these results show that density does not merely act as a static structural parameter, but 
serves as a control parameter for dynamical properties. 
In particular, changing density systematically modifies lattice stiffness, which in turn alters the energy landscape experienced by carriers, leading to changes in anharmonicity, cooperative dynamics, and the spatial structure of mobility.
This establishes a direct mechanistic link between density and the dynamical quantities emphasized in this work-phonon anharmonicity, collective diffusion, and heterogeneous carrier dynamics-rather than a purely structural interpretation.
In real materials, similar effects would manifest not through a single density parameter, but via changes in lattice volume, bonding strength, or local coordination environments induced by composition or external tuning. While such modifications inevitably involve changes in material chemistry, they effectively tune the same physical quantities identified in our model, namely lattice softness and dynamical heterogeneity, which govern transport behavior.

\subsection{Connection between collective hopping, migration entropy, and effective activation barriers}
\begin{figure*}[!htbp]
    \centering
    \includegraphics[width=\linewidth]{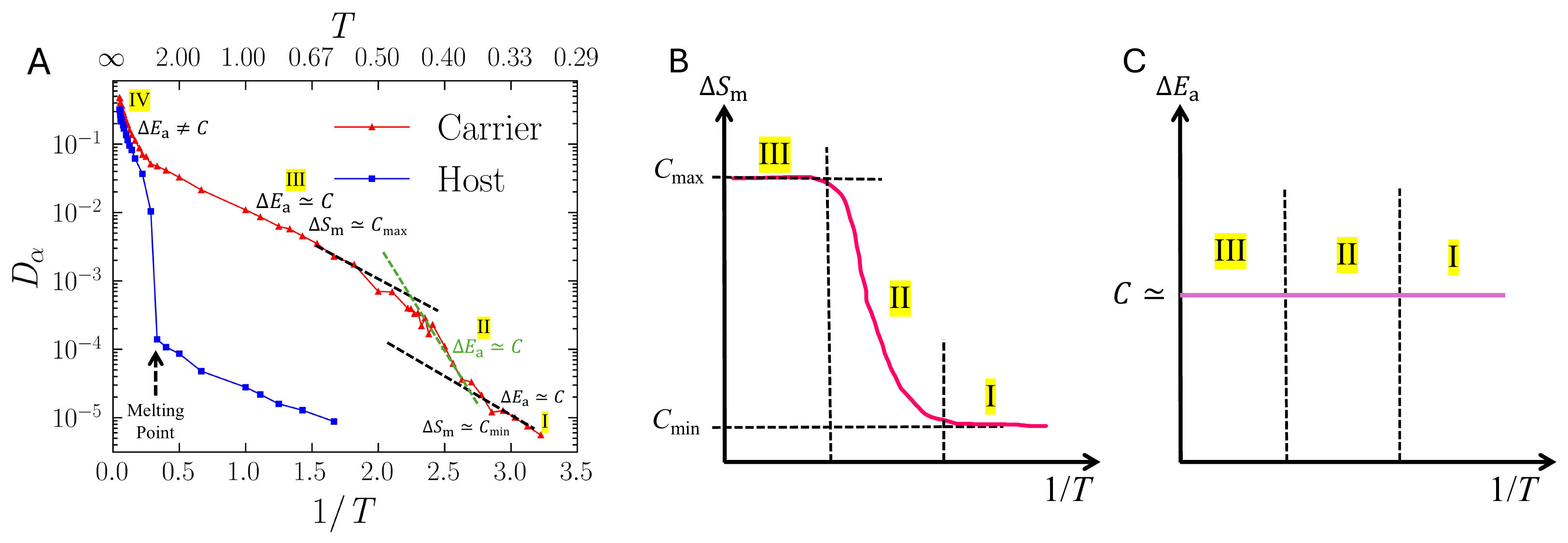}
   \caption{Effective energy barrier of the system. (\textit{A}) Arrhenius plot ($D_\alpha$ vs.\ $1/T$), analogous to Fig.~2\textit{A} (of the main text), with annotations highlighting the behavior of the activation energy $\Delta E_\mathrm{a}$ and migration entropy $\Delta S_\mathrm{m}$ across temperature regimes. (\textit{B}) Schematic illustration of the migration entropy $\Delta S_\mathrm{m}$ in each regime, emphasizing its qualitative variation and its role in governing transport. (\textit{C}) Activation energy $\Delta E_\mathrm{a}$ as a function of $1/T$, showing an approximately constant behavior in regimes I--III, consistent with an intact host lattice.}
    \label{fig:effective-barrier-entropy}
\end{figure*}
To clarify the relationship between collective hopping dynamics and activation barriers, it is important to distinguish between the activation energy $\Delta E_\mathrm{a}$ and the effective free-energy barrier $\Delta F$ governing ionic transport. Within a transition-state framework, the diffusivity can be expressed as \cite{LiJACS2023, KrauskopfChemMater2017}

\begin{align}
    D \sim \exp\left(\frac{\Delta S_\mathrm{m}}{k_\mathrm{B}}\right)\exp\left(-\frac{\Delta E_\mathrm{a}}{k_\mathrm{B} T}\right) = \exp\left(-\frac{\Delta F}{k_\mathrm{B} T}\right),
    \label{diffusivity-energy-entropy}
\end{align}
where $\Delta S_\mathrm{m}$ is the migration entropy, and
\begin{align}
\Delta F = \Delta E_\mathrm{a} - T \Delta S_\mathrm{m},
\label{effective-energy-barrier}
\end{align}
is the free-energy barrier (effective activation energy). From the Boltzmann principle, the migration entropy can be interpreted as
\begin{align}
\Delta S_\mathrm{m} = k_\mathrm{B} \ln\left(\frac{\Omega^\ddagger}{\Omega_0}\right),
\label{migration-entropy}
\end{align}
where $\Omega^\ddagger$ and $\Omega_0$ denote the number of accessible configurations at the transition state and in the initial state, respectively.

As shown in Fig.~\ref{fig:effective-barrier-entropy}, $\Delta S_\mathrm{m}$ increases from $C_{\rm min}$ to $C_{\rm max}$ as the system evolves from regime I $\to$ II $\to$ III (sublattice melting). Here, $C_{\rm min} \sim O(1)$ corresponds to a limited number of localized hopping carriers, whereas $C_{\rm max} \sim O(N)$ reflects a regime in which a macroscopic number of carriers contribute to transport. Since the energetic contribution $\Delta E_\mathrm{a}$ remains approximately constant due to the persistence of the host lattice, the increase in $\Delta S_\mathrm{m}$ leads to a reduction of the effective free-energy barrier $\Delta F$, thereby explaining the observed temperature dependence of $D$.

On the basis of this framework, the behavior observed in our system (Fig.~\ref{fig:effective-barrier-entropy}) can be understood as follows:

\noindent {\bf Region I:} (low temperature):  
Transport is dominated by single-particle hopping. The number of accessible pathways is limited, leading to an approximately constant migration entropy $\Delta S_\mathrm{m}$. As a result, the diffusivity follows standard Arrhenius behavior.\\
\noindent {\bf Region II:} (onset of sublattice melting):  
This regime is characterized by the emergence of collective hopping events. These cooperative motions significantly increase the number of accessible configurations, leading to a strong temperature dependence of $\Delta S_\mathrm{m}$. This entropic enhancement produces a deviation from Arrhenius behavior and gives rise to an apparent change in slope.\\
\noindent {\bf Region III:} (fully developed superionic state):  
Although the migration entropy is substantially larger than in Region I, it becomes approximately temperature-independent due to the homogeneous availability of multiple diffusion pathways. Consequently, Arrhenius behavior is recovered, with a slope similar to Region I. \\
\noindent {\bf Region IV:}  
Complete melting, including the host lattice. In this regime, $\Delta E_\mathrm{a}$ differs from that in the crystalline host phase.
\begin{figure*}[!htbp]
    \centering
    \includegraphics[width=\linewidth]{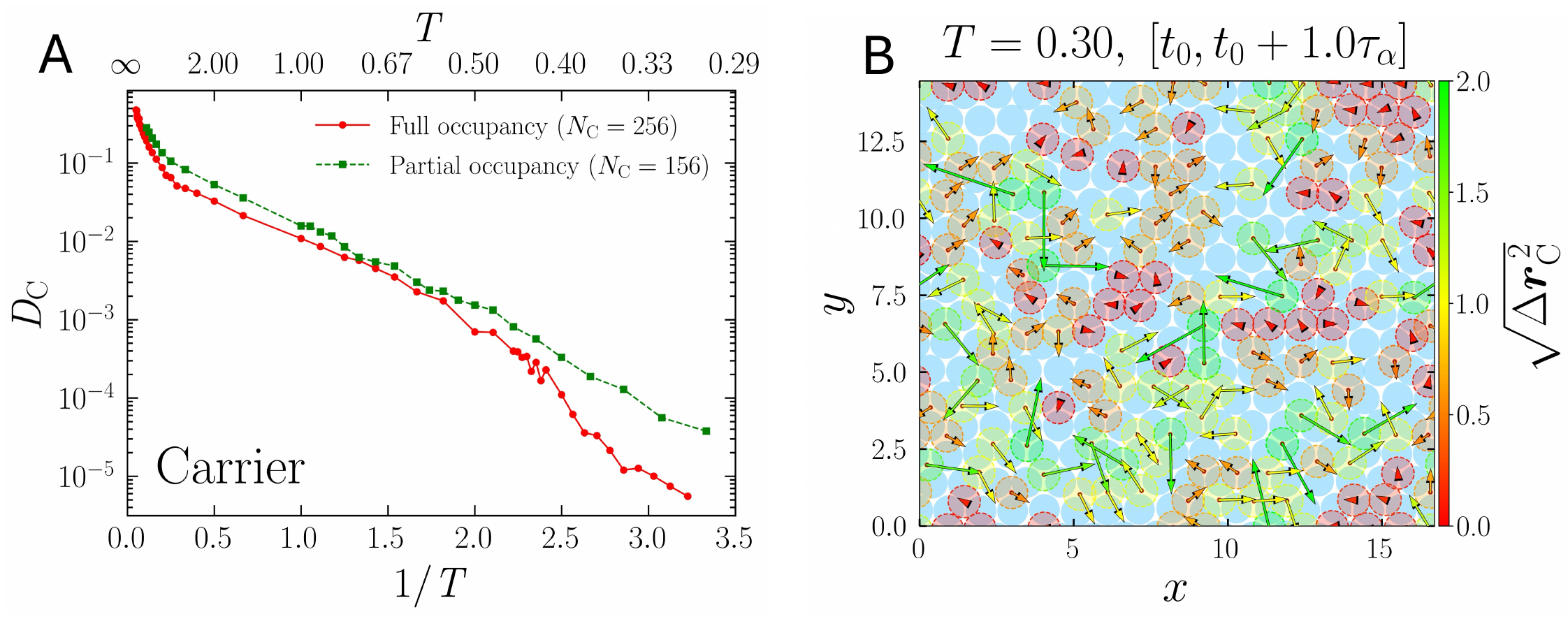}
    \caption{Dynamics of partially occupied carrier systems. (\textit{A}) Arrhenius plot of the diffusion coefficient for carriers ($D_\mathrm{C}$) as a function of inverse temperature ($1/T$), comparing a fully occupied system (same as for our 2D NAP model for $\phi = 0.85$) with a partially occupied system. While both systems exhibit Arrhenius-like behavior at high temperatures, the partially occupied system shows higher diffusivity and weaker dynamical slowing at low temperatures due to reduced crowding and vacancy-assisted motion. (\textit{B}) Spatial mobility map of the partially occupied carrier system at $T=0.30$. Arrows indicate carrier displacements over the observation interval $\left [ t_0,\, t_0 + \,\tau_\alpha \right ]$. Unlike the fully occupied case, the dynamics do not exhibit pronounced string-like cooperative motion or strongly localized dynamical heterogeneity, indicating that transport is dominated primarily by spatially dispersed single-particle hopping events under partial occupancy.}
    \label{fig:partial-occumancy}
\end{figure*}
\subsection{Effect of partial occupancy on carrier dynamics in our NAP model}
From a physical perspective, reducing the occupancy increases the availability of vacant sites (see Fig.~\ref{fig:partial-occumancy}), which can weaken the degree of cooperativity in hopping dynamics. This may shift the onset of collective dynamics to different temperatures or reduce the magnitude of the migration entropy enhancement. Nevertheless, the central mechanism identified in this work-namely, the increase of migration entropy $\Delta S_\mathrm{m}$ due to the proliferation of accessible hopping pathways, leading to a reduction in the effective barrier $\Delta F$-is expected to remain valid (see previous section as well). Thus, our framework generalizes from collective transport in dense systems to vacancy-mediated transport in dilute systems, providing a unified description across different occupancy regimes.

To further examine this effect, we considered a reduced carrier concentration ($N_\mathrm{C}:N_\mathrm{H}=156:256$) and compared the resulting dynamics with the fully occupied case ($N_\mathrm{C}:N_\mathrm{H}=256:256$), as shown in Fig.~\ref{fig:partial-occumancy}\textit{A}. At high temperatures, both systems exhibit comparable Arrhenius-like transport behavior. However, clear differences emerge in the low-temperature regime, where the partially occupied system maintains systematically higher diffusivity and exhibits a substantially weaker dynamical slowdown than the fully occupied case. In contrast to the pronounced crossover behavior observed for full occupancy, the partially occupied system displays an approximately single-Arrhenius temperature dependence over the investigated range.

To clarify the microscopic origin of this behavior, we analyzed the spatial mobility patterns in the low-temperature diffusive regime ($T=0.30$), shown in Fig.~\ref{fig:partial-occumancy}\textit{B}. Unlike the fully occupied system, the partially occupied state does not exhibit pronounced string-like cooperative motion or strongly heterogeneous collective rearrangements. Instead, the dynamics are dominated by relatively isolated and spatially dispersed hopping events, indicating that diffusion primarily proceeds through vacancy-assisted single-particle motion. The suppression of collective hopping pathways under partial occupancy is therefore consistent with the disappearance of the crossover between cooperative and noncooperative transport regimes, leading to the observed single-Arrhenius behavior.
\begin{figure*}[!htbp]
\centering
\includegraphics[width=\linewidth]{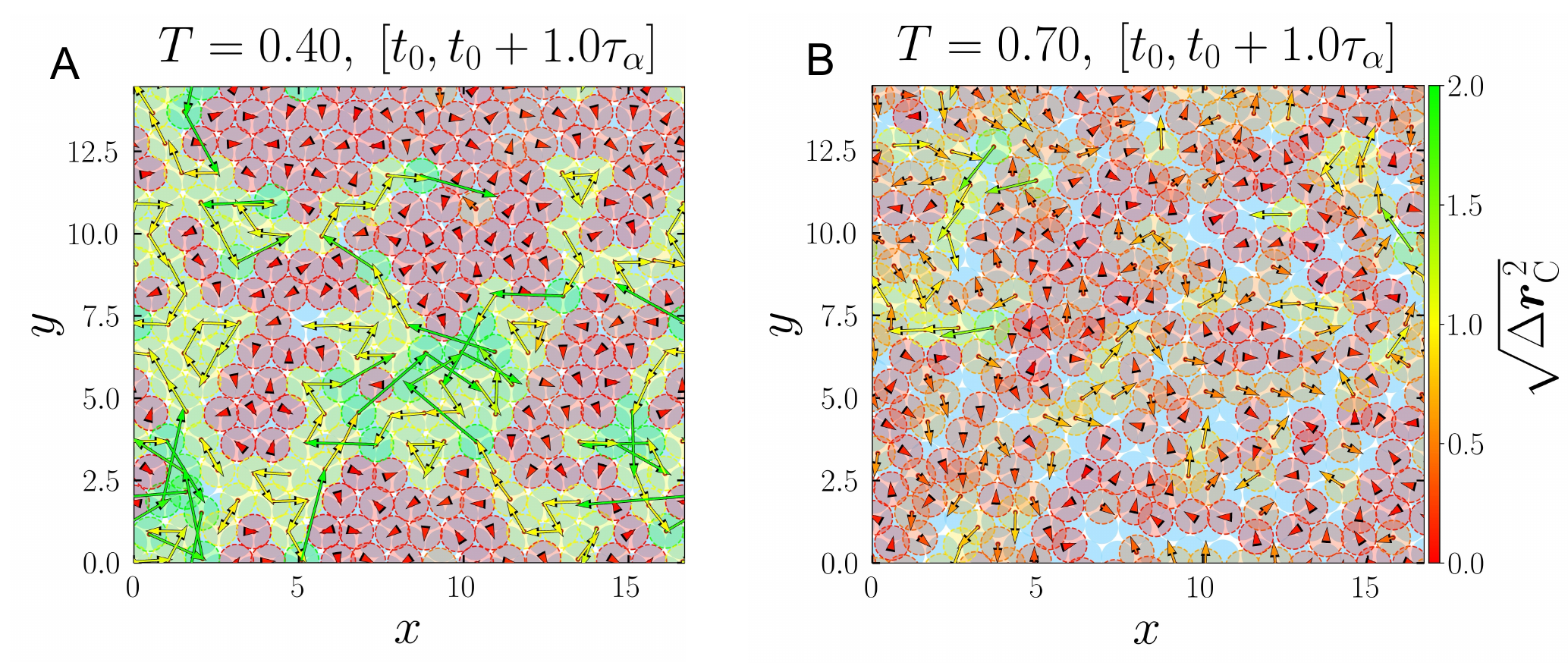}
\caption{Temperature evolution of spatial mobility maps at $\phi=0.85$. (\textit{A}) $T=0.4$ ($T<T_\mathrm{f}$) and (\textit{B}) $T=0.7$ ($T>T_\mathrm{f}$). At low temperature, pronounced dynamical heterogeneity is observed through coexisting mobile and localized regions, which diminishes as the system transitions to a more homogeneous dynamical state at higher temperatures.}
\label{fig:spatialmap}
\end{figure*}
\subsection{Temperature dependence of spatial dynamical heterogeneity}

To directly visualize the evolution of dynamical heterogeneity, we examine spatial maps of particle mobility across different temperature regimes. Figure~\ref{fig:spatialmap} shows the distribution of particle displacements over the interval $[t_0, t_0+\tau_\alpha]$ at representative temperatures below and above the sublattice melting temperature $T_\mathrm{f}$.

At low temperatures ($T<T_\mathrm{f}$), the system exhibits pronounced spatial heterogeneity, characterized by the coexistence of highly mobile regions and localized, quasi-crystalline domains. As the temperature increases toward $T_\mathrm{f}$, these heterogeneous regions progressively diminish, indicating a reduction in spatio-temporal correlations in particle motion. Above the sublattice melting regime ($T>T_\mathrm{f}$), the dynamics become increasingly homogeneous, with particle mobility distributed more uniformly across the system.

\newcounter{movie}

\renewcommand{\figurename}{Movie}
\setcounter{figure}{0}

\renewcommand{\thefigure}{S\arabic{figure}}
\stepcounter{movie}
\subsection{Unconstrained carrier transport in the fully molten regime ($T=7.0$)}
\begin{figure}[t!]
    \centering
    \includegraphics[width=0.9\linewidth]{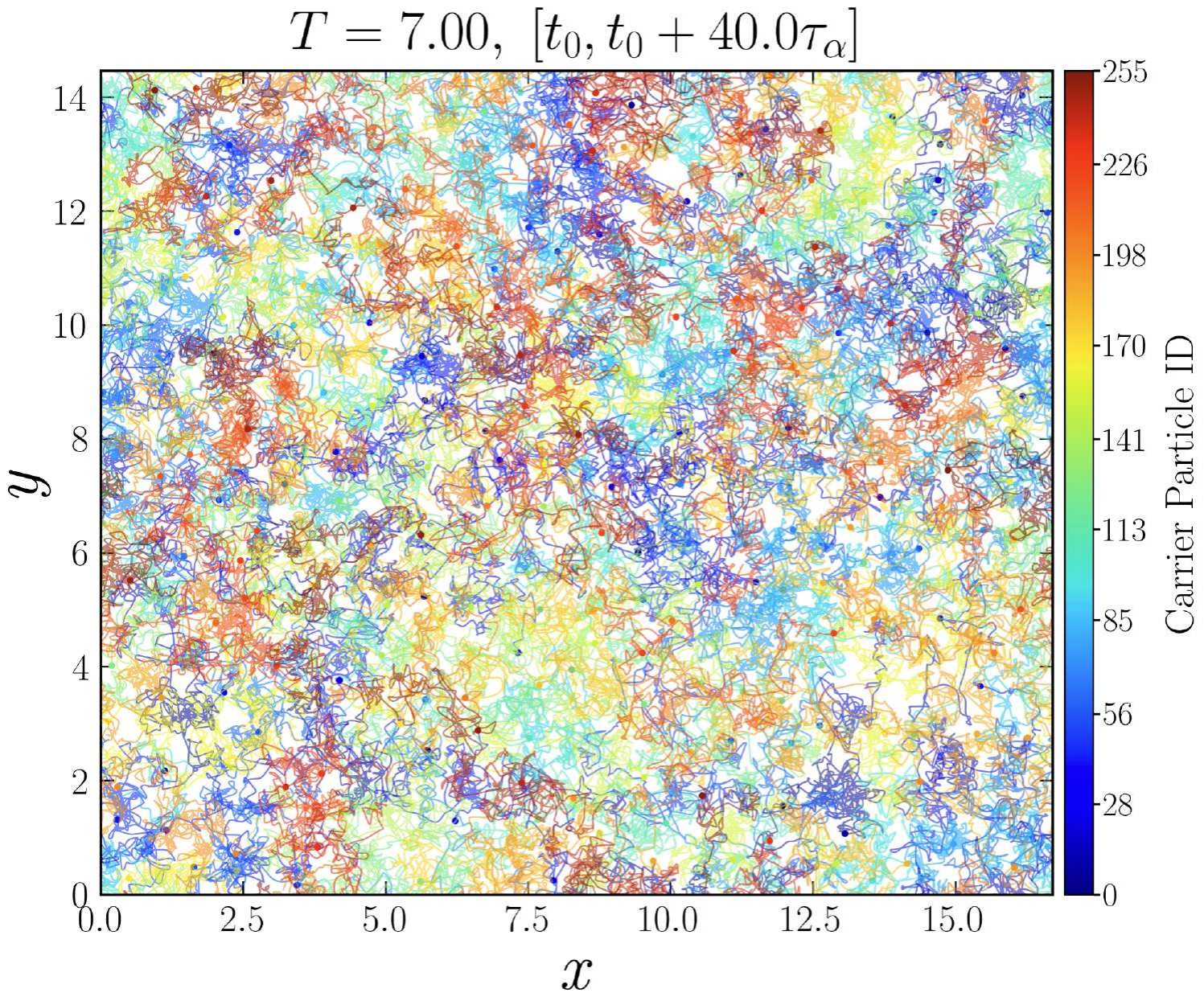}

    \caption{Carrier trajectories at $T=7.0$ show homogeneous, liquid-like motion due to melting of both carrier and host sublattices. Movie link is here: https://doi.org/10.5281/zenodo.20482338.}
    \label{movie:s1}
\end{figure}
To visualize carrier transport within the host sublattice, we generated trajectory movies for a two-dimensional system at area packing fraction $\phi = 0.85$. Three movies were prepared to illustrate carrier dynamics across different thermal regimes. Particle identities are color-coded to track individual trajectories and assess the spatial homogeneity of carrier motion throughout the system.
Movie \ref{movie:s1} corresponds to a high-temperature state ($T = 7.0$), well above the melting temperature. The trajectories are shown over a time window $\left [ t_0,\, t_0 + 40\,\tau_\alpha \right ]$, where $\tau_\alpha$ denotes the structural relaxation time at $T = 7.0$. In this regime, both the carrier and host sublattices are fully molten, eliminating geometric constraints imposed by the host. As a result, no bottleneck effects are observed, and the system exhibits liquid-like behavior characterized by homogeneous carrier motion and uniform spatial distribution.

\subsection{Sublattice melting with liquid-like carriers and rigid host lattice ($T=2.5$)}
\stepcounter{movie}
\begin{figure}[t!]
    \centering
    \includegraphics[width=0.9\linewidth]{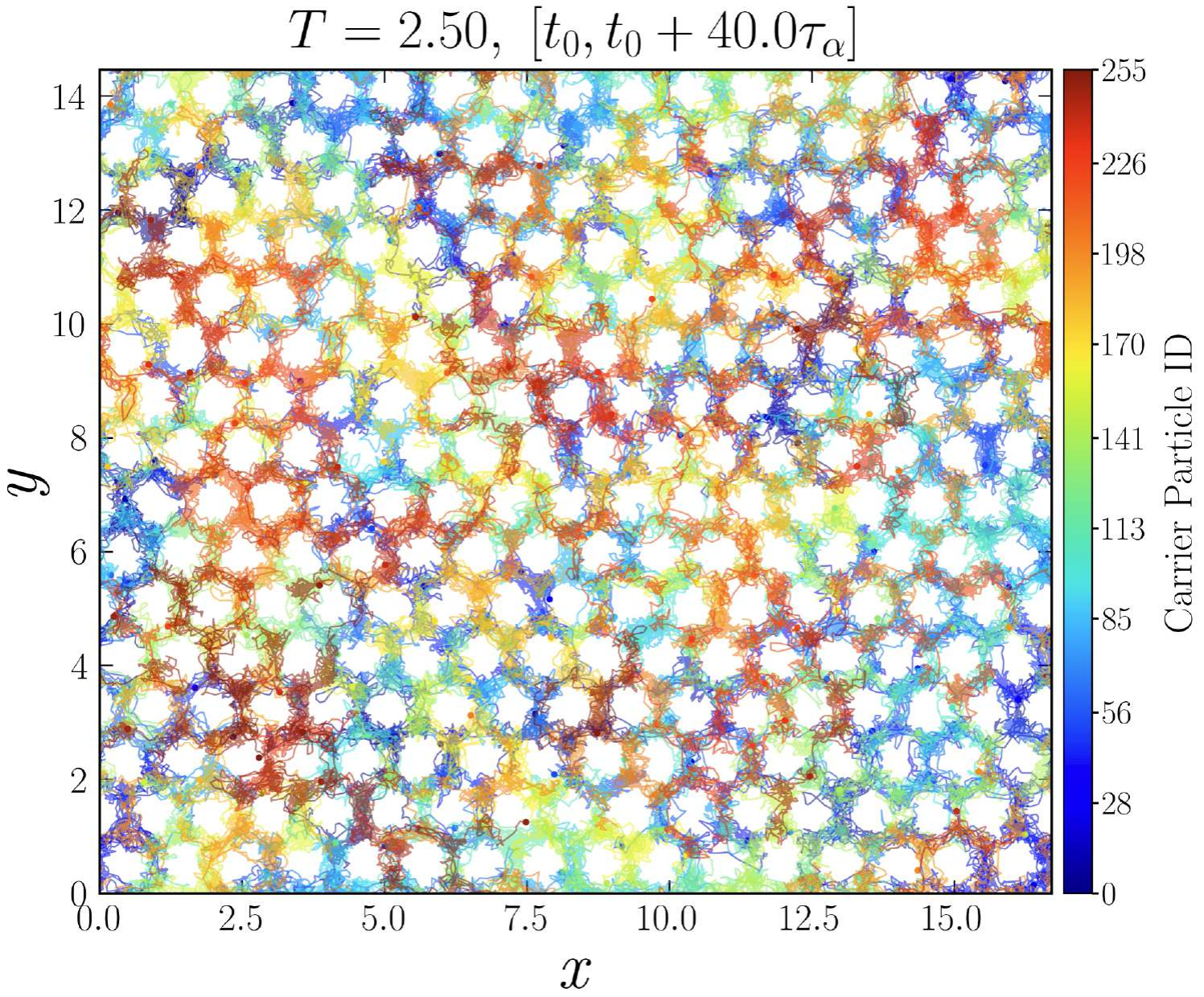}

    \caption{At $T=2.5$, the carrier sublattice is molten while the host lattice remains rigid, leading to bottleneck-guided but system-spanning transport. Movie link is here: https://doi.org/10.5281/zenodo.20482338.}
    \label{movie:s2}
\end{figure}
Movie \ref{movie:s2} shows particle trajectories at temperature $T = 2.5$ for a two-dimensional system at area packing fraction $\phi = 0.85$. The trajectories are recorded over the time window $\left [t_0,\, t_0 + 40\,\tau_\alpha \right]$, where $\tau_\alpha$ is the structural relaxation time at T = 2.5. The system is initialized from a well-relaxed, spatially homogeneous configuration, indicating that it is in a steady state.

Despite the relatively high temperature, the dynamics reveal a clear separation between carrier and host degrees of freedom. While the carrier particles exhibit liquid-like motion and undergo long-range transport, the host lattice remains largely immobile over the observation window. As a result, the carrier trajectories develop pronounced bottleneck structures around the fixed host sites, leading to persistent low-occupancy regions coinciding with the host lattice positions.
This coexistence of a mobile carrier sublattice with a structurally stable host lattice provides direct dynamical evidence of sublattice melting: the carrier sublattice is fully melted and percolates through the system, whereas the host lattice retains its positional order. The resulting heterogeneous flow pathways highlight the constrained nature of carrier transport imposed by the rigid host framework, even at elevated temperatures.

\bibliography{soft}

\end{document}